\documentclass{aa}

   \usepackage{graphicx}
   \usepackage{txfonts}

\begin{document}

\title{VLBI observations of nineteen GHz-Peaked-Spectrum radio sources at 1.6 GHz}
   \author{X. Liu
          \inst{1}
           \and
           L. Cui\inst{1,2}
           \and
           W. -F. Luo\inst{1,2}
           \and
           W. -Z. Shi\inst{1,2}
           \and
           H. -G. Song \inst{1
           }
 }

   \offprints{X. Liu: liux@ms.xjb.ac.cn}

   \institute{National Astronomical Observatories/Urumqi Observatory, CAS,
40-5 South Beijing Road, Urumqi 830011, China\\
         \and Graduate University of the Chinese Academy of Sciences, Beijing 100049, China\\
      }

   \date{Received  / Accepted}

   \abstract{Aims and Methods: We present the results of VLBI observations of
nineteen GHz-Peaked-Spectrum (GPS) radio sources at 1.6~GHz. Of
them, 15 sources are selected from the Parkes Half Jansky (PHJ)
sample (Snellen 2002), 4 others are from our previous observation
list. We aimed at imaging the structure of GPS sources, searching
for Compact Symmetric Objects (CSOs) and studying the absorption
for the convex radio spectra of GPS sources.

Results: We obtained total intensity 1.6~GHz VLBI images of 17
sources for the first time. Of them, 80\% show mini-double-lobe
radio structure, indicating that they are CSOs or candidates, and
their host AGNs could be edge-on to us. This result suggests that
there is a high incidence of mini double-lobe sources (or CSOs) in
the PHJ sample. The sources J0323+0534, J1135$-$0021, J1352+0232,
J2058+0540, J2123$-$0112 and J2325$-$0344 with measured redshift,
showing double-lobe structure with sizes of $<1$ kpc, are
classified as CSOs. Three sources J1057+0012, J1600$-$0037 and
J1753+2750 are considered as core-jet sources according to their
morphologies and flux variability.

}

\maketitle

   \titlerunning{EVN observations of nineteen GHz-Peaked-Spectrum (GPS) radio sources at 1.6 GHz}
   \authorrunning{X. Liu et al.}


\section{Introduction}

GHz-Peaked-Spectrum (GPS) radio sources are powerful
($P_{\rm1.4~GHz}\geq\rm10^{25}\,W\,Hz^{-1}$), compact ($\leq
1$~kpc), and have convex radio spectra, and they make up a
significant fraction ($\approx$ 10\%) of the bright radio source
sample, see O'Dea (1998) for a review. In general, the presence of
large scale emission associated with GPS galaxies is rare, about a
few percent in a GPS sample (Stanghellini et al. 2005). Most GPS
sources appear to be truly compact and isolated.

Their small size is most likely due to their youth ($<{10^4}$
years) according to a spectral aging analysis (Murgia 2003). A
couple of GPS sources are certainly young radio sources whose
kinematic age from lobe proper motions has been measured and these
sources are also identified as Compact Symmetric Objects (CSOs).
There is compelling evidence in favour of the youth scenario of
GPS sources and CSOs, see e.g. Owsianik \& Conway (1998), Tschager
et al. (2000), Polatidis \& Conway (2003), and Orienti et al.
(2007). The GPS sources and CSOs are the key objects to study the
early evolution of powerful radio-loud AGN. A unification scenario
assumes that GPS sources evolve into Compact Steep Spectrum
sources (1-15 kpc), which in turn, evolve into classical extended
radio sources ($> 15$~kpc), i.e. FR I/II radio sources (Fanti et
al. 1995, Snellen et al. 2000, de Vries et al. 2007).

GPS $\it{galaxies}$ are dominated by lobe/jet emission on both
sides of the central engine, and are thought to be relatively free
of beaming effects.
The GPS galaxies show very low polarization (about less than 0.5\%
at 5~GHz, Dallacasa 2004, Xiang et al. 2006). The low integrated
polarization could be due to large Faraday depths around the radio
source, which would depolarize the radio emission, implying that
their host-AGNs are probably edge-on to us.

Since GPS sources live in the narrow line region of AGN, it is
likely that their low frequency radio emission will be absorbed
due to either synchrotron self-absorption or free-free absorption,
giving rise to a peaked radio spectrum. Therefore, GPS sources are
also suitable for studying radio absorption and scattering in
AGNs.

We have carried out EVN (European VLBI Network) observations of 19
GPS sources, 15 of them are from the Parkes Half Jansky (PHJ)
sample (Snellen et al. 2002) with declination $>-5^\circ$ and not
observed with VLBI before. Four sources are from our previous
observation list which we have observed with the EVN at 2.3/8.4
GHz and/or 5 GHz (see Xiang et al. 2005, 2006). We aimed at
imaging the GPS sources at 1.6~GHz, in order to confirm whether
the GPS sources are double-lobe sources, and to find CSO
candidates. For the sources with observations at 2.3, 5.0 and 8.4
GHz, the 1.6 GHz images will further provide information on their
source structure and intensity at lower frequency, for further
spectral study of the GPS sources in the future.

\section{Observations and data reduction}

The observations were carried out on 3 March 2006 at 1.65~GHz
using the MK5 recording system with a bandwidth of 32 MHz and
sample rate of 256 Mbps in dual circular polarization. The EVN
antennae in this experiment were Effelsberg, Westerbork, Jodrell,
Medicina, Noto, Onsala, Torun, Hartebeesthoek, Urumqi and
Shanghai. Snapshot observations of 19 sources
(Table~\ref{table:gps}) in a total of 24 hours were made. OQ208
and DA193 were observed as calibrators. The data correlation was
completed at JIVE.

The total flux densities of the sources were also measured at 5
GHz with Urumqi 25m telescope in order to find any flux
variability. The values are listed in Table~\ref{table:flux}.

The Astronomical Image Processing System (AIPS) has been used for
editing, a-priori calibration, fringe-fitting, self-calibration,
imaging and model fitting of the data.

\section{Results and comments on individual sources}

We list the basic information of the sources in
Table~\ref{table:gps}, and the parameters derived from the VLBI
images in Table~\ref{table:vlbi}. We comment on the results of
each source and give a short discussion. We use $S \propto
\nu^{-\alpha}$ to define the spectral index. Optical information
and redshifts of the GPS sources in the PHJ sample are given by
de Vries et al. (2007), as listed in Table~\ref{table:gps}.

   \begin{table*}
         \caption[]{The GPS sources. Columns (1),(2) source names; (3) optical identification (G: galaxy, QSO: quasar, EF: empty field); (4) optical
         magnitude; (5)
         redshift (de Vries et al. 2007, those with * are a photometric estimated by Tinti et al. 2005); (6) linear scale factor pc/mas
         [$H_{0}=71 km s^{-1} Mpc^{-1}$ and $q_{0}=0.5$ have been
         assumed]; (7) maximum angular size from the observation; (8) maximum linear size; (9) 1.4 GHz flux density from the
         NVSS; (10)
         2.7 GHz flux density from Snellen sample and the NED; (11) low frequency spectral index; (12) higher frequency spectral index (
         computed from columns 9 and 10); (13) turnover frequency; (14)
peak flux density; (15) references for the spectral information, 1
Snellen et al. 2002, 2 de Vries et al. 1997, 3 Stanghellini et al.
1998, where $S \propto \nu^{-\alpha}$.}
         $$
\begin{tabular}{ccccccccccccccc}
            \hline
            \hline
            \noalign{\smallskip}
            1&2&3&4&5&6&7&8&9&10&11&12&13&14&15\\
$Source$ && $id$ & $m_{R}$ & $z$ & $pc/mas$ & $\theta$ & $L$ & $S_{1.4}$ & $S_{2.7}$ & $\alpha_{l}$ & $\alpha_{h}$ & $\nu_{m}$ & $S_{m}$ & $ref$ \\
& & & & && mas & pc & mJy &Jy & & & GHz & Jy & \\
            \noalign{\smallskip}
            \hline
            \noalign{\smallskip}

J0210+0419 &B0208+040& G   & $18.3Ks$  & 1.5* &6.1    & 90  &          & 948 &0.56 &  &0.80  & 0.4 & 1.3 & 1 \\

J0323+0534 &4C+05.14& G   & 19.2    & 0.1785 &2.7   &180 &     490 &  2793 &1.60 &  &0.85   & 0.4 & 7.1 & 1 \\

J0433$-$0229 & 4C$-$02.17&G  & 19.1  & 0.530 &  5.1  & 80 & 408          & 1462 &1.04  &  &0.52    & 0.4 & 3.0 & 1 \\

J0913+1454 & B0910+151&G  & $22.9$ & 0.47* & 4.9   & 80  &            & 881 &0.54  &  &0.75  & 0.6 & 1.1& 1 \\

J1057+0012& B1054+004&G   & $22.3$ & 0.65* & 5.5  & 80? &            & 898 &0.58  &     &0.67  & 0.4   & 1.6  & 1\\

J1109+1043& B1107+109&G   & 22.6   & 0.55*  & 5.2   & 60  &            & 1481 &0.80  &  & 0.94  & 0.5 & 2.4& 1 \\

J1135$-$0021 &4C$-$00.45& G   & 21.9  & 0.975 & 6.0 & 120 &  720          & 1268 &0.76  &   &0.78   & 0.4 &2.9 &1  \\

J1203+0414& B1200+045& QSO   & 18.8  & 1.221 &  6.1     & 75 &   458      & 1146 &0.85  &  &0.45   & 0.4 &1.4 &1  \\

J1352+0232 &B1349+027& G  & 20.0  &  0.607  & 5.4   & 170  & 918           &1145 &0.78  &  &0.58   & 0.4 & 2.0 & 1 \\

J1352+1107 & 4C+11.46&G   & $21.0$ & 0.891&  5.9 & 50 &   295         & 1538 &0.78  &  &1.03   & 0.4 & 3.6 & 1 \\

J1600$-$0037 & B1557$-$004 &G  &        &      &        &  50 &            & 1168 &0.54    &  &1.17   & 1.0 & 1.2 & 1 \\

J1648+0242 &4C+02.43& G   & $22.1$  & 0.824&  5.8  &    &           &      & 0.61 &   &   &0.4  &3.4 & 1 \\

J2058+0540 & 4C+05.78&G   & 23.4  & 1.381 & 6.1   &160  &     970    & 1213 &0.65  &   &0.95   & 0.4 &3.1  & 1 \\

J2123$-$0112 & B2121$-$014 & G   & 23.3  & 1.158 & 6.1 & 80     &   488     & 1087 &0.64 & -0.56 & 0.75  & 0.5 & 1.8 & 2 \\

J2325$-$0344 &B2322$-$040 & G  & 23.5 &  1.509  &  6.0     & 75  &  450          & 1224  & 0.91  & -0.42 & 0.75  & 1.4 & 1.3 & 2 \\

J0917+1113 & B0914+114 & EF   &      &       &             & 190 &            & 800   &0.31  & -0.1  & 1.6   & 0.3 & 2.3 & 3 \\

J1753+2750 & B1751+278 & G   & 21.7  & 0.86* & 5.9        & 50   &              &  625  & 0.46 & -0.27 & 0.57  & 1.4 &0.6 &2 \\

J1826+2708 & B1824+271 & G  & 22.9  &       &             & 45  &            &  332   & 0.23 & -0.39 & 0.75  & 1.0 & 0.4 & 2 \\

J2325+7917 & B2323+790 & G   & 19.5V &       &             & 32  &            & 1136   &     &-0.3  &0.75   &1.4  &1.2 & 2 \\

            \noalign{\smallskip}
            \hline
        \end{tabular}{}
         \label{table:gps}
         $$
   \end{table*}

   \begin{table*}
         \caption[]{Source flux and possible variability, columns 2-4 are flux densities at 5.0 GHz (PKS90),
         4.85 GHz (Gregory \& Condon 1991, and Griffith et al. 1995) and
         4.85 GHz flux measured with the Urumqi 25m telescope on 2007/1/24 (J1648+0242, J2058+0540, 1824+271 and
         2121$-$014 were not well measured due to source confusion or weak);
         column 5 is a flux variability computed from columns 3,4.
}
         $$
         \begin{tabular}{ccccc}
            \hline
            \hline
            \noalign{\smallskip}
$Source$ & $S_{5.0}$ & $S_{4.85}$ & $S_{4.85Ur}$ & $\delta S_{4.85}$  \\

& mJy & mJy & mJy & \%  \\
            \noalign{\smallskip}
            \hline
            \noalign{\smallskip}

J0210+0419 &  300  &298$\pm19$& $302\pm10$& 1.3 $\pm$3.1      \\

J0323+0534 &  830&819$\pm$44&$868\pm9$&  6.0$\pm$4.6 \\

J0433$-$0229 & 640 &640$\pm$35&$637\pm14$& -0.5$\pm$3.3  \\

J0913+1454 & 300 &315$\pm$43&$297\pm8$& -5.7$\pm$10.3  \\

J1057+0012& 370&396$\pm$23 &$351\pm6$& -11.4$\pm$3.6 \\

J1109+1043& 400 &408$\pm$56&$428\pm8$&  4.9$\pm$12.4 \\

J1135$-$0021 &440  &446$\pm$25  & $427\pm8$    &  -4.3$\pm$3.6    \\

J1203+0414&  520& 640$\pm$35  & $611\pm7$ &  -4.5 $\pm$4.1\\

J1352+0232 & 470 &       & $469\pm7$ &   \\

J1352+1107 & 410& 447$\pm$62  & $418\pm5$ & -6.5$\pm$11.9  \\

J1600$-$0037 & 180   & 187$\pm$14& $212\pm3$   & 13.4$\pm$6.9    \\

J1648+0242 &260&  337$\pm$20    &   &    \\

J2058+0540 & 340 & 356$\pm$21  &    &    \\

2121$-$014 & 320 & 345$\pm$21 && \\

2322$-$040 & 500 &524$\pm$29 &  $545\pm20$  &  4.0 $\pm$1.9 \\

0914+114 & 140    &134$\pm$19 &  $140\pm1$ &   4.5 $\pm$14.1    \\

1751+278 &  & 298$\pm$39& $292\pm6$& -2.0$\pm$10.8 \\

1824+271 &      & 122$\pm$17& &          \\

2323+790 &     &  & $491\pm7$&       \\


OQ208    & &2421$\pm$217 & $2514\pm12$& 3.8$\pm$8.8\\

           \noalign{\smallskip}
            \hline
           \end{tabular}{}
         $$
         \label{table:flux}
   \end{table*}

   \begin{table*}
         \caption[]{The component parameters of the VLBI images at 1.6 GHz. The columns give: (1)
         source name and possible classification (CSOc: CSO candidate, cj: core-jet); (2)
         total cleaned flux density of image at 1.6 GHz; (3)
         component identification labled to Xiang et al. 2002, 2005, 2006; (4),(5) peak and
         integral intensity of a fitted Gaussian component at 1.6 GHz in the AIPS task JMFIT; (6),(7) major/minor axes and position angle
         of component at 1.6 GHz; (8),(9) distance and position angle relative to the first component; (10) brightness temperature
         of component.}
         $$
\begin{tabular}{cccccccccc}
          \hline
           \hline
            \noalign{\smallskip}
    1&2&3&4&5&6&7&8&9&10\\
$Name$      &$S_{vlbi}$  &$Comp$ &  $Sp$ &  $Sint$  & $\theta_{1}\times\theta_{2}$ & $PA$     & $d$                 &    $PA$ & $Tb$\\
class       & mJy        &       &  mJy  &   mJy    &     mas$\times$mas           & $^\circ$ &  mas                &    $^\circ$  &$10^{8}K^{\circ}$\\
            \noalign{\smallskip}
            \hline
            \noalign{\smallskip}

J0210+0419   &715       &A       &290    &375       & 5.7 $\times$ 2.9               &169     &0                    &                      &11.9              \\
   CSOc          &          &B       &118    &205       &  10 $\times$ 3.6               &175     &68.2  $\pm$ 0.1      &-153.2 $\pm$ 0.1    &2.2               \\
J0323+0534   &1497      &A       &536    &1270      &32.6 $\times$ 11.1              &70      &0                    &                      &0.5               \\
   CSO          &          &B       &117     &548       &57.5 $\times$ 21.7              &18     &122.7 $\pm$ 2.9      &-167.2 $\pm$ 0.4    &0.1               \\
J0433$-$0229 &1095      &A       &407    &1045       &  14.5 $\times$ 4.7                 &171     &0                    &                      &2.4               \\
    CSOc/cj         &          &B       &49     &111       & 9.9 $\times$ 5.5               &88      & 53.0 $\pm$ 0.3      &   164.4 $\pm$ 0.2    &0.3               \\
J0913+1454   &796       &A       &211    &501       & 8.2 $\times$ 4.9               &65      &0                    &                      &2.1               \\
  CSOc           &          &B       &55     &157       & 8.9 $\times$ 6.2               &80      & 56.8 $\pm$ 0.1      &    73.3 $\pm$ 0.1    &0.4               \\
J1057+0012   &810       &A       &381    &550       & 3.6 $\times$ 2.7               &6.6     &0                    &                      &17.5              \\
    cj         &          &B       &37     &58        & 6.2 $\times$ 2.1               &9.7     &  9.7 $\pm$ 0.2      &   114.8 $\pm$ 0.9    &1.3               \\
J1109+1043   &1370      &A       &420    &984       & 5.7 $\times$ 4.7               &110     &0                    &                      &6.6               \\
  CSOc           &          &B       &143    &311       & 5.4 $\times$ 4.2               &91      & 46.3 $\pm$ 0.1      &   104.4 $\pm$ 0.1    &2.6               \\
J1135$-$0021 &1025       &A       &279    &454       & 6.6 $\times$ 2.7               &142     &0                    &                      &4.9               \\
   CSO         &          &B       &143    &271       & 8.6 $\times$ 3.0               &153     & 85.8 $\pm$ 0.1      &   164.4 $\pm$ 0.1    &1.7               \\
J1203+0414   &1029      &A       &571    &850       & 4.7 $\times$ 2.9               &107     &0                    &                      &25.1              \\
  CSOc     &          &B       &51     &81        & 5.0 $\times$ 3.8               &78      & 18.1 $\pm$ 0.2      &   103.0 $\pm$ 0.4    &1.6               \\
             &          &C       &31     &35        &   6 $\times$ 6                 &0       & 58.2 $\pm$ 0.2      &   104.2 $\pm$ 0.2    &1.9               \\
J1352+0232   &885      &A       &173    &480       & 7.4 $\times$ 4.5               &53      &0                    &                      &2.1               \\
  CSO           &          &B       &30     &120       & 8.9 $\times$ 6.3               &110     &165.7 $\pm$ 0.3      &-111.9 $\pm$ 0.1    &0.2               \\
J1352+1107   &896       &A       &198    &395       & 8.1 $\times$ 5.4               &11.7    &0                    &                      &2.0               \\
  CSOc/cj           &          &B       &106    &202       & 6.8 $\times$ 6.3               &10      &  3.5 $\pm$ 0.1      &    53.1 $\pm$ 0.1    &1.1               \\
J1600$-$0037 &936       &A       &394    &607       & 5.4 $\times$ 4.1               &157     &0                    &                      &                  \\
 cj     &          &B       &125    &255       & 7.9 $\times$ 5.0               &86      & 24.7 $\pm$ 0.1      &    83.5 $\pm$ 0.1    &                  \\
J2058+0540   &914       &A       &356    &513       &   7.6 $\times$ 3.7                 &163     &0                    &                      &8.2               \\
   CSO          &          &B       &182    &403      &  12.6 $\times$ 4.5                 &128     &127.2 $\pm$ 0.1      &   172.1 $\pm$ 0.1    &2.3               \\
2121$-$014   &976       &A       &363    &594       & 5.1 $\times$ 3.9               &123      &0                    &                      &10.4              \\
 CSO         &          &C       &191    &415       & 7.4 $\times$ 5.0               &126     & 59.4 $\pm$ 0.1      &    85.4 $\pm$ 0.1    &2.9               \\
2322$-$040   &965       &A       &229    &509       &  14.6 $\times$ 3.3                 &161     &0                    &                      &3.6               \\
CSO          &          &B       &74     &160       &  11.7 $\times$ 5.3                 &162     & 40.6 $\pm$ 0.2      &   171.3 $\pm$ 0.1    &0.8               \\
             &          &C       &65     &196       &  19.3 $\times$ 4.7                 &1     & 22.0 $\pm$ 0.4      &   171.1 $\pm$ 0.2    &0.5               \\
0914+114     &578       &A       &37     &50        & 4.9 $\times$ 2.7               &16      &0                    &                      &            \\
CSO          &          &B       &29     &65        & 9.1 $\times$ 4.3               &66      & 45.6 $\pm$ 0.1      &    80.1 $\pm$ 0.1    &                 \\
             &          &C       &242    &360       & 4.2 $\times$ 3.9               &90      & 84.2 $\pm$ 0.1      &    81.6 $\pm$ 0.1    &                 \\
         &          &E       &28     &40        & 4.4 $\times$ 3.4               &63      & 86.1 $\pm$ 0.1      & -95.9 $\pm$ 0.1    &                  \\
1751+278     &596       &A       &400    &522       & 4.8 $\times$ 2.9               &71      &0                    &                      &14.5              \\
 cj       &          &B       &26     &37        &   8 $\times$ 3                 &15      & 20.5 $\pm$ 0.1      &-128.6 $\pm$ 0.2    &0.5               \\
             &          &C       &12.4   &21        &   7 $\times$ 5                 &59      & 26.6 $\pm$ 0.3      &-119.8 $\pm$ 0.3    &0.2               \\
         &          &D       &8      &20        &19.5 $\times$ 3.6               &176     & 41.4 $\pm$ 0.4      &-104.4 $\pm$ 0.8    &0.1               \\
1824+271     &296       &A       &145    &174       & 3.9 $\times$ 2.1               &164     &0                    &                      &                 \\
 CSO         &          &B       &42     &69        & 6.1 $\times$ 4.3               &137     & 21.8 $\pm$ 0.1      & -83.4 $\pm$ 0.1    &                 \\
2323+790     &900       &A       &439   &621       & 5.8 $\times$ 2.1               &159     &0                    &                      &                  \\
 CSOc        &          &B+C       &107    &155     & 7.9 $\times$ 3.1               &118     & 19.2 $\pm$ 0.1      & -71.1 $\pm$ 0.1    &                  \\

\noalign{\smallskip}
            \hline
            \end{tabular}{}
             \label{table:vlbi}
             $$


\end{table*}

\subsection{J0210+0419 (PKS B0208+040)}

\begin{figure}
     \includegraphics[width=7cm]{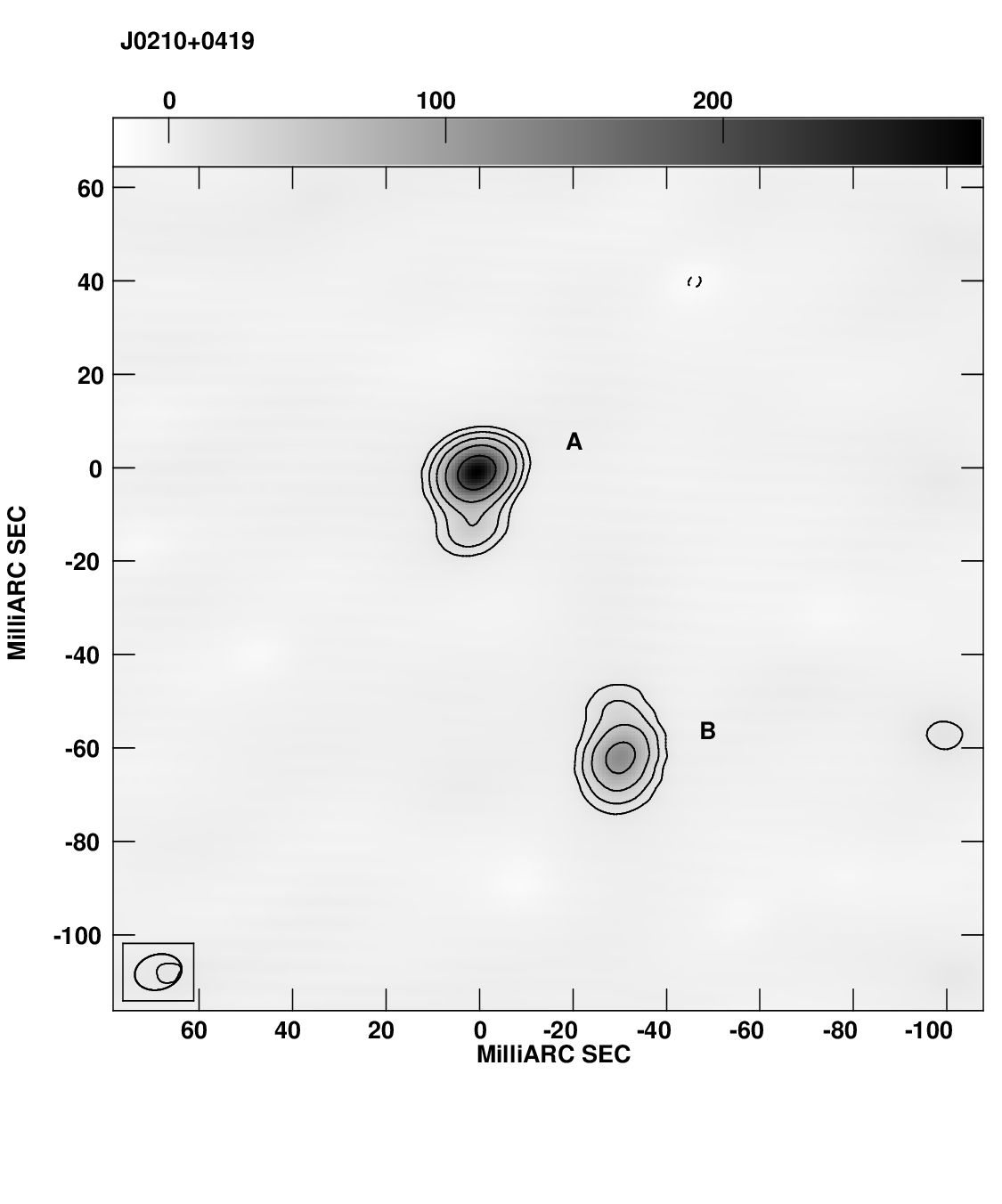}
     \caption{J0210+0419 at 1.65 GHz, the restoring beam is
        $10.2\times7.5$ mas with PA $-75.5^{\circ}$, the peak is 293 mJy/beam, the contours
        are 12 mJy/beam times levels -1, 1, 2, 4, 8, 16, 32, 64, 100, 200, 400, 800,
        and the same levels are used in the following images, the grey scale unit is mJy.}
      \label{fig1}
   \end{figure}

The 1.6 GHz VLBI image (Fig.~\ref{fig1}) is the first VLBI image
of the source. It shows a double-lobe structure, and is most
likely a CSO. Optical observations did not result in an
identification with a lower limit of $m_{R}>24.1$, but it is
identified with a magnitude of $Ks$=18.3 (de Vries et al. 2007).

\subsection{J0323+0534 (4C+05.14)}

\begin{figure}
     \includegraphics[width=7cm]{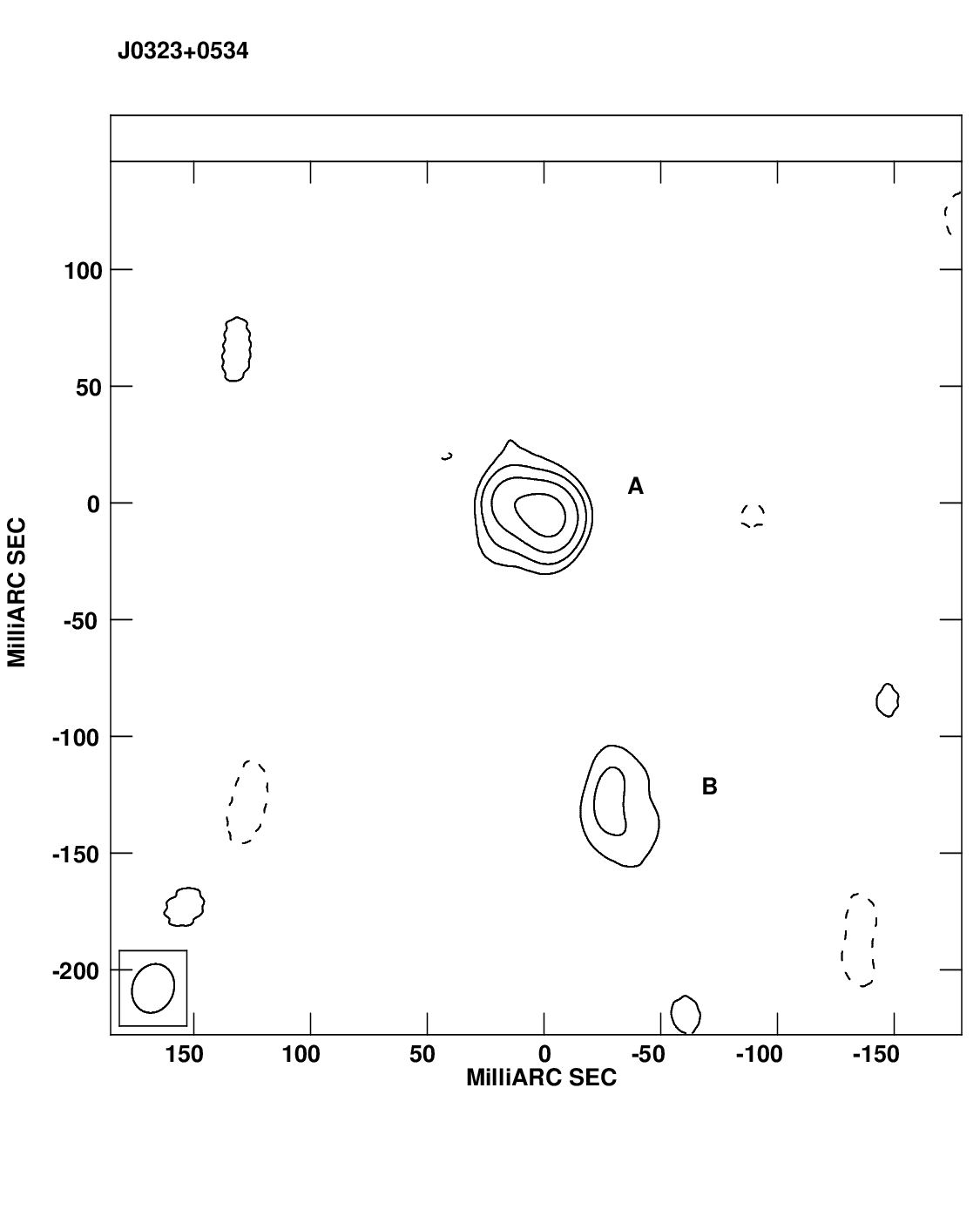}
     \caption{J0323+0534 at 1.65 GHz, the restoring beam is
        $21.3\times17.7$ mas with PA $-20.2^{\circ}$, the peak is 586 mJy/beam, the first contour
        is 50 mJy/beam.}
      \label{fig2}
   \end{figure}

The 1.6 GHz VLBI image (Fig.~\ref{fig2}) is the first VLBI image
of the source, and it exhibits a strong diffuse component and a
weak extended component in the south. Both are likely lobe
emission. About 38\% total flux density (estimated from
Table~\ref{table:gps}) is resolved out in the VLBI image, due to
the diffuse components. For its size of 490 pc, the source can be
a CSO.

\subsection{J0433$-$0229 (4C$-$02.17)}

\begin{figure}
     \includegraphics[width=7cm]{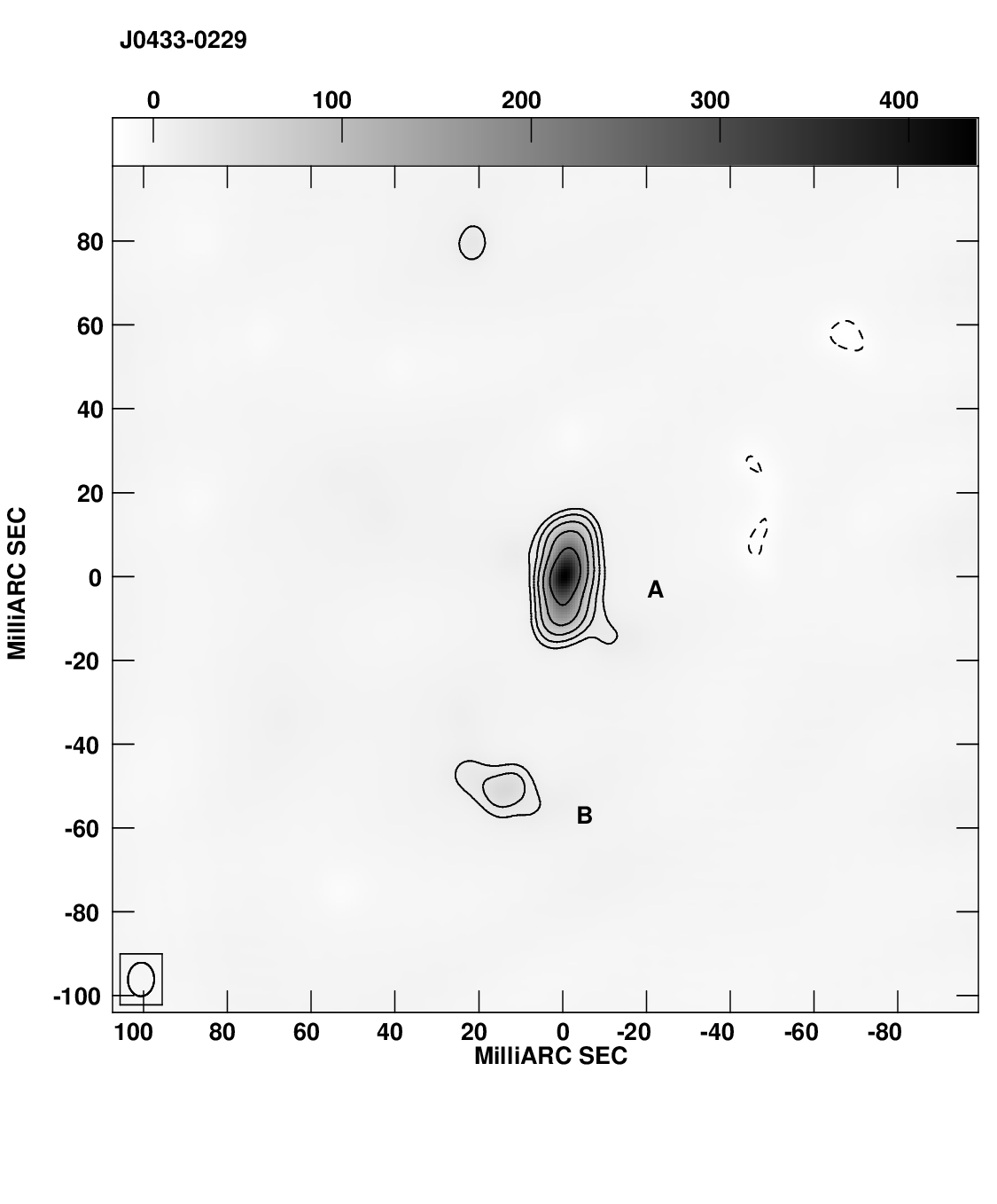}
     \caption{J0433$-$0229 at 1.65 GHz, the restoring beam is
        $8.0\times6.3$ mas with PA $-1.7^{\circ}$, the peak is 435 mJy/beam, the first contour
        is 15 mJy/beam.}
      \label{fig3}
   \end{figure}

The 1.6 GHz VLBI image (Fig.~\ref{fig3}) is the first VLBI image
of the source, and the main component is diffuse and extended in
the north-south direction, and a possible weak component in the
south. About 18\% total flux density (estimated from
Table~\ref{table:gps}) is resolved out in the VLBI image. Either a
core-jet or a CSO classification is possible for the source.

\subsection{J0913+1454 (PKS B0910+151)}

\begin{figure}
     \includegraphics[width=7cm]{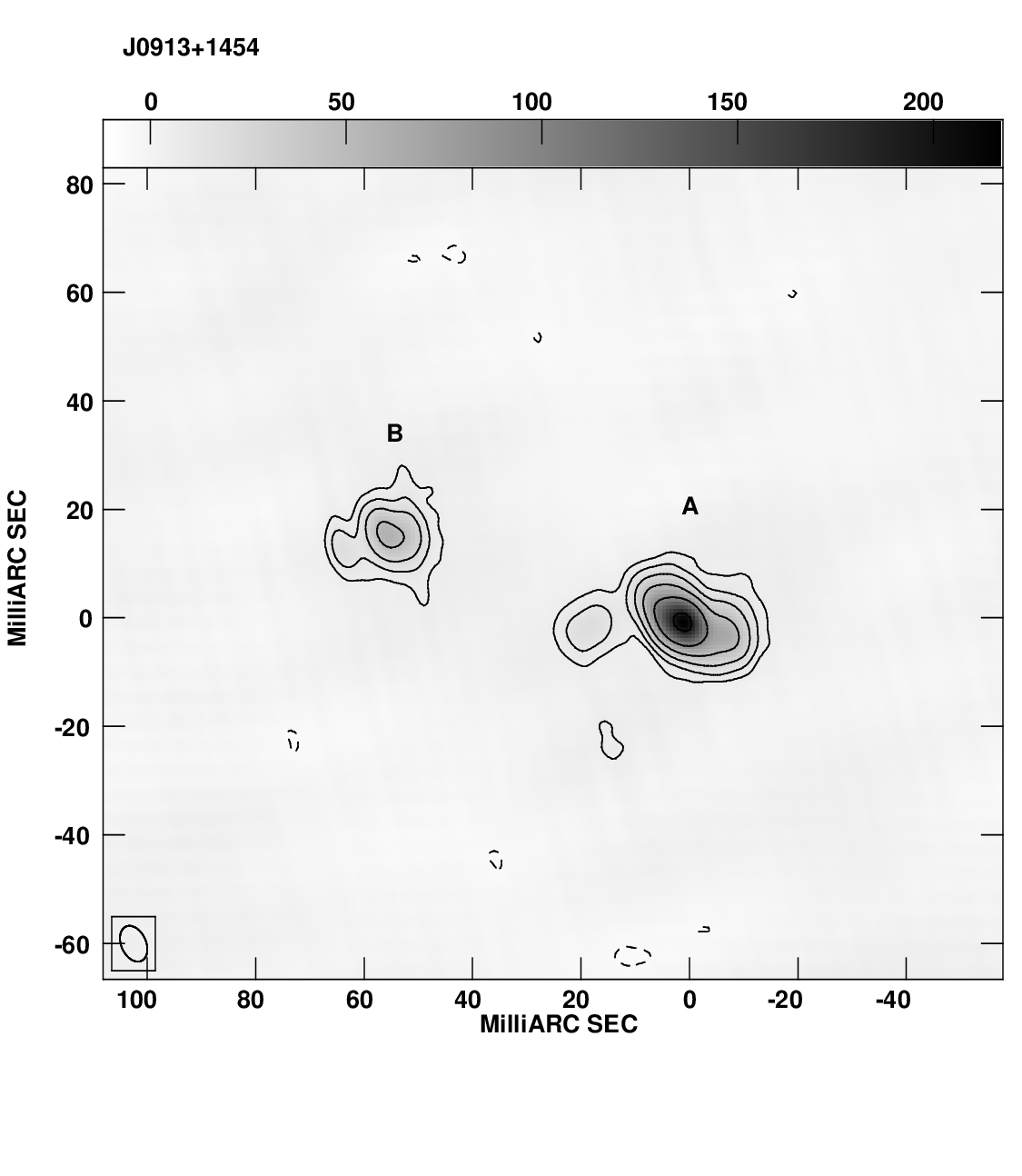}
     \caption{J0913+1454 at 1.65 GHz, the restoring beam is
        $6.9\times4.6$ mas with PA $24.4^{\circ}$, the peak is 217 mJy/beam, the first contour
        is 6 mJy/beam.}
      \label{fig4}
   \end{figure}

The 1.6 GHz VLBI image (Fig.~\ref{fig4}) is the first VLBI image
of the source. It shows double structure and both components are
further resolved. There is probably a hot-spot imbedded in the
bright one. We consider it as a CSO candidate.

\subsection{J1057+0012 (PKS B1054+004)}

\begin{figure}
     \includegraphics[width=7cm]{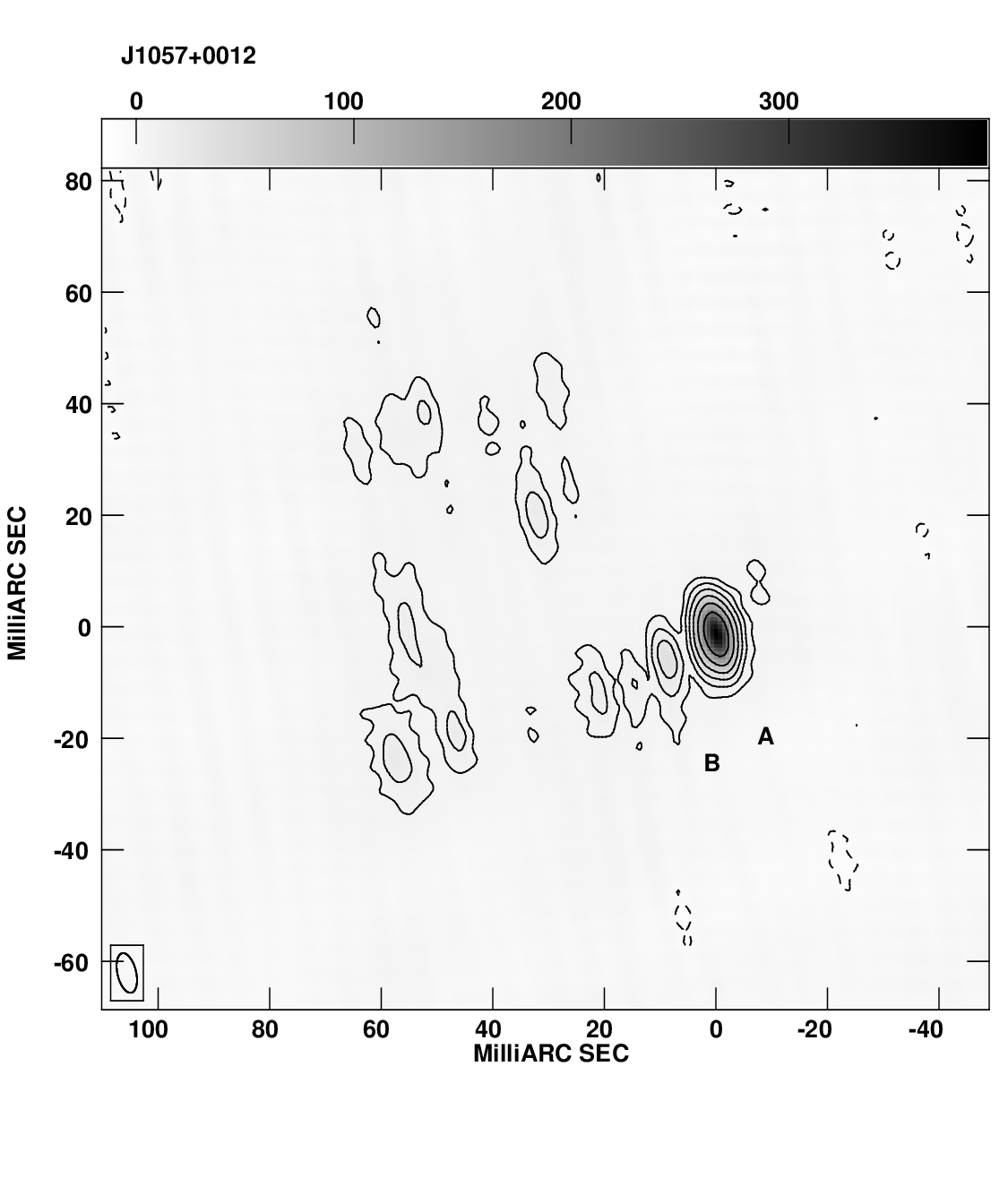}
     \caption{J1057+0012 at 1.65 GHz, the restoring beam is
        $7.3\times3.3$ mas with PA $13.4^{\circ}$, the peak is 390 mJy/beam, the first contour
        is 6 mJy/beam.}
      \label{fig5}
   \end{figure}

The 1.6 GHz VLBI image (Fig.~\ref{fig5}) is the first VLBI image
of the source. There is a bright compact component followed by a
secondary component and a series of possible weak components in
the east, indicating this is a core-jet source. A flux variability
of $(-11.4\pm3.6)\%$ over 15 years at 5 GHz, as reported in
Table~\ref{table:flux}, is consistent with the core-jet
classification.

\subsection{J1109+1043 (PKS B1107+109)}

\begin{figure}
     \includegraphics[width=7cm]{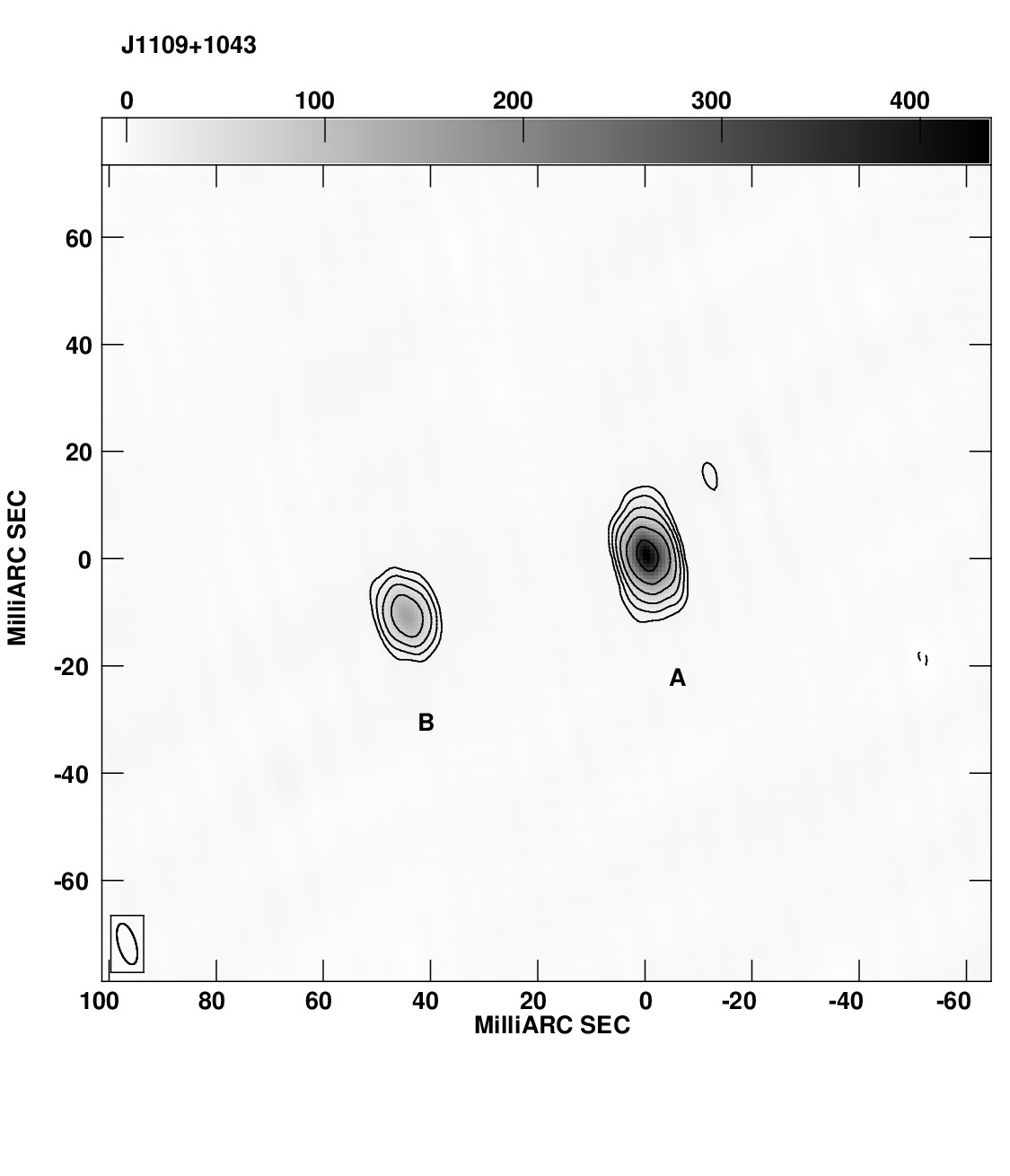}
     \caption{J1109+1043 at 1.65 GHz, the restoring beam is
        $7.9\times3.3$ mas with PA $16.5^{\circ}$, the peak is 434 mJy/beam, the first contour
        is 10 mJy/beam.}
      \label{fig6}
   \end{figure}

The 1.6 GHz VLBI image (Fig.~\ref{fig6}) is the first VLBI image
of the source. It is a double structure, and can be a CSO
candidate. The total flux density (1270 mJy estimated from
Table~\ref{table:gps}) is completely restored in the VLBI image
(1370 mJy, increased by 8\%). There is also an indication of total
flux increasing $(4.9\pm12.4)\%$ at 5 GHz in
Table~\ref{table:flux} but with a large error.

\subsection{J1135$-$0021 (4C$-$00.45)}

\begin{figure}
     \includegraphics[width=7cm]{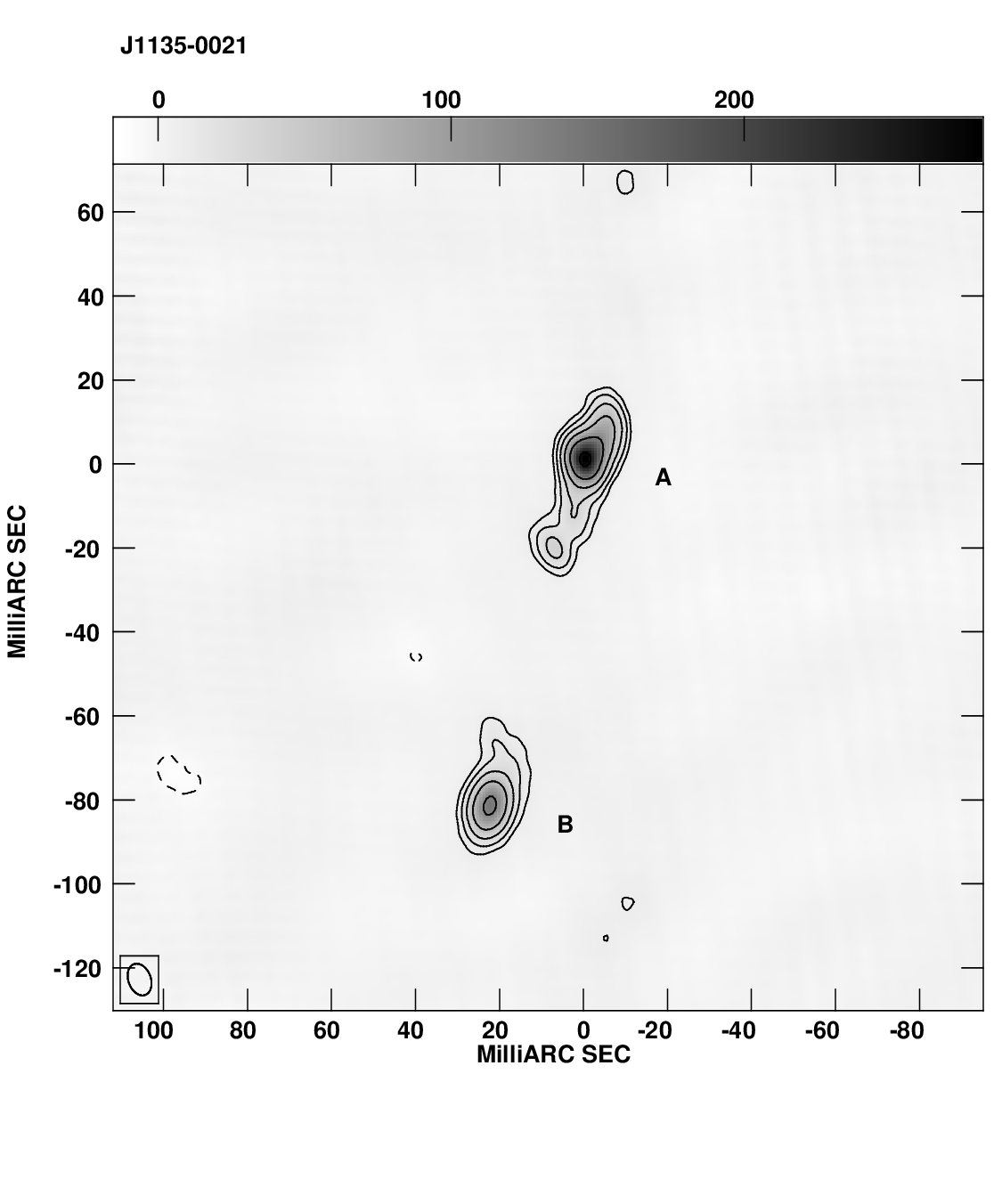}
     \caption{J1135$-$0021 at 1.65 GHz, the restoring beam is
        $7.8\times5.3$ mas with PA $20.4^{\circ}$, the peak is 281 mJy/beam, the first contour
        is 8 mJy/beam.}
      \label{fig7}
   \end{figure}

The 1.6 GHz VLBI image (Fig.~\ref{fig7}) is the first VLBI image
of the source. It shows a double-lobe structure, and with a size
of 720 pc, we classify the source as a CSO.

\subsection{J1203+0414 (PKS B1200+045)}

\begin{figure}
     \includegraphics[width=7cm]{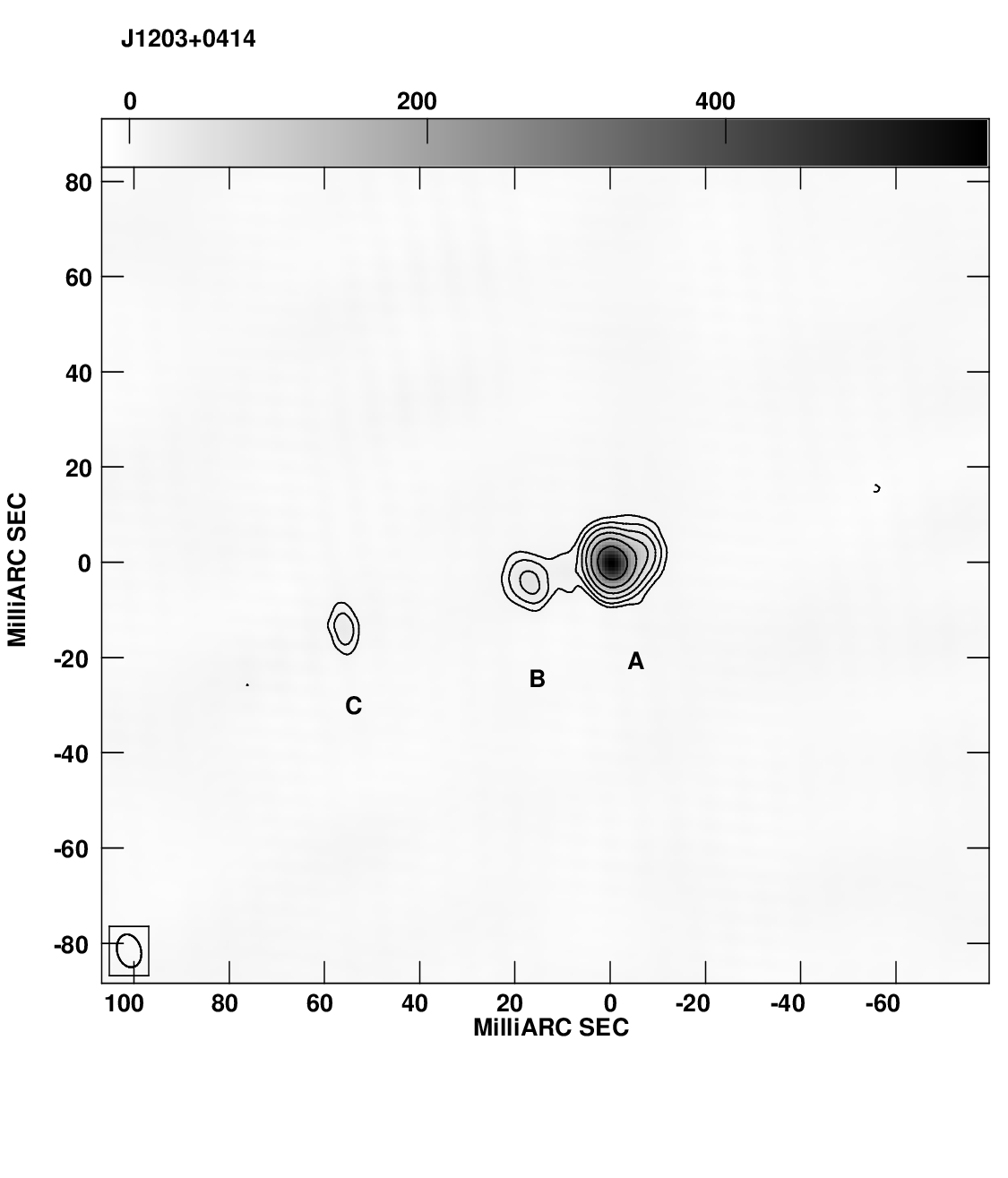}
     \caption{J1203+0414 at 1.65 GHz, the restoring beam is
        $7.0\times4.9$ mas with PA $15.2^{\circ}$, the peak is 575 mJy/beam, the first contour
        is 10 mJy/beam.}
      \label{fig8}
   \end{figure}

The 1.6 GHz VLBI image (Fig.~\ref{fig8}) is the first VLBI image
of the source. The triple structure may consist of a core and two
sided emission, or a one sided core-jet source. The quasar as
newly identified by de Vries et al (2007), is possibly a core-jet
one, but still we keep the source as a CSO candidate.

\subsection{J1352+0232 (PKS B1349+027)}

\begin{figure}
     \includegraphics[width=7cm]{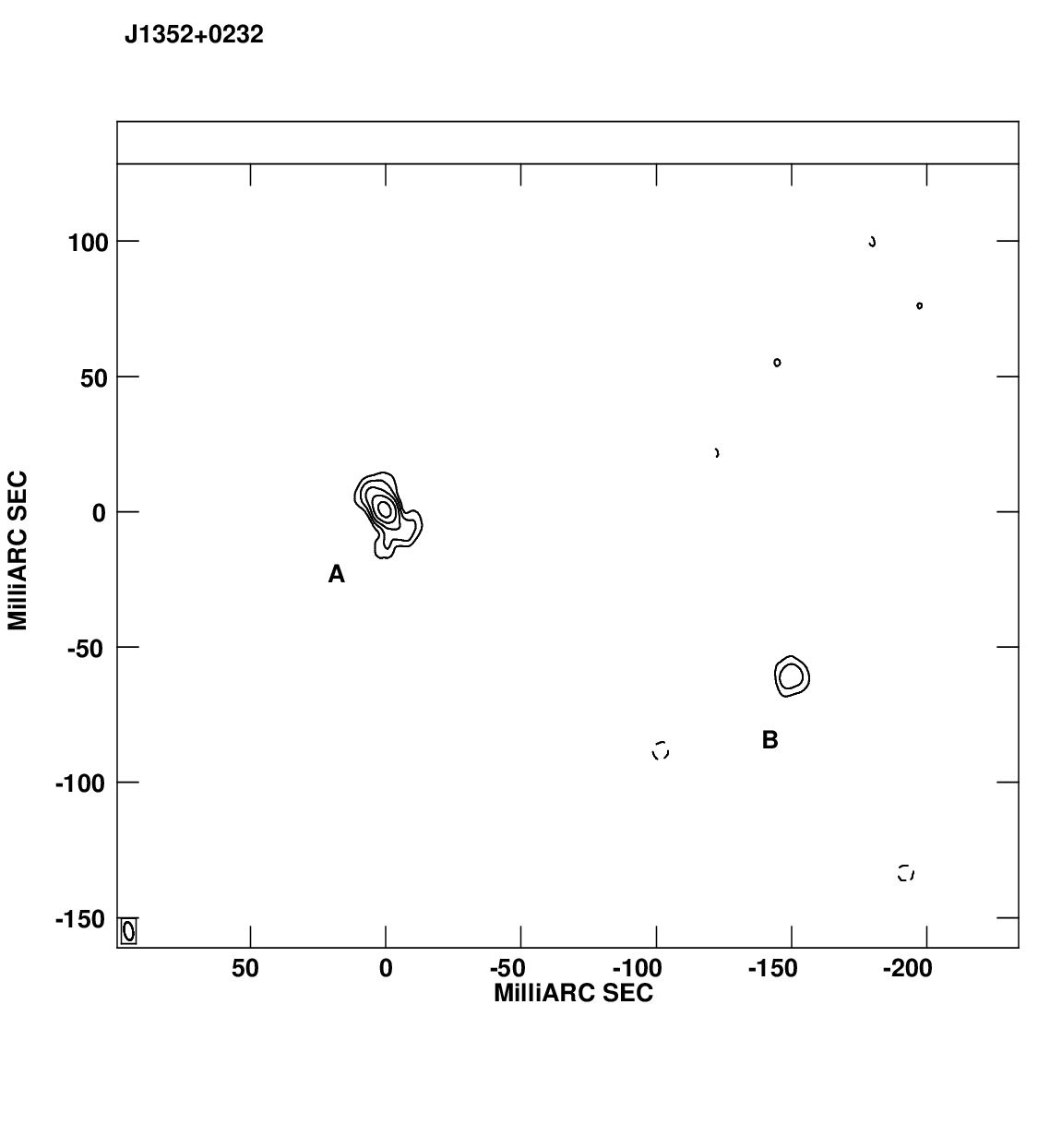}
     \caption{J1352+0232 at 1.65 GHz, the restoring beam is
        $6.7\times3.2$ mas with PA $9.9^{\circ}$, the peak is 183 mJy/beam, the first contour
        is 8 mJy/beam.}
      \label{fig9}
   \end{figure}

The 1.6 GHz VLBI image (Fig.~\ref{fig9}) is the first VLBI image
of the source. It shows a double-lobe like structure, for its size
of 918 pc we consider it as a CSO.

\subsection{J1352+1107 (4C+11.46)}

\begin{figure}
     \includegraphics[width=7cm]{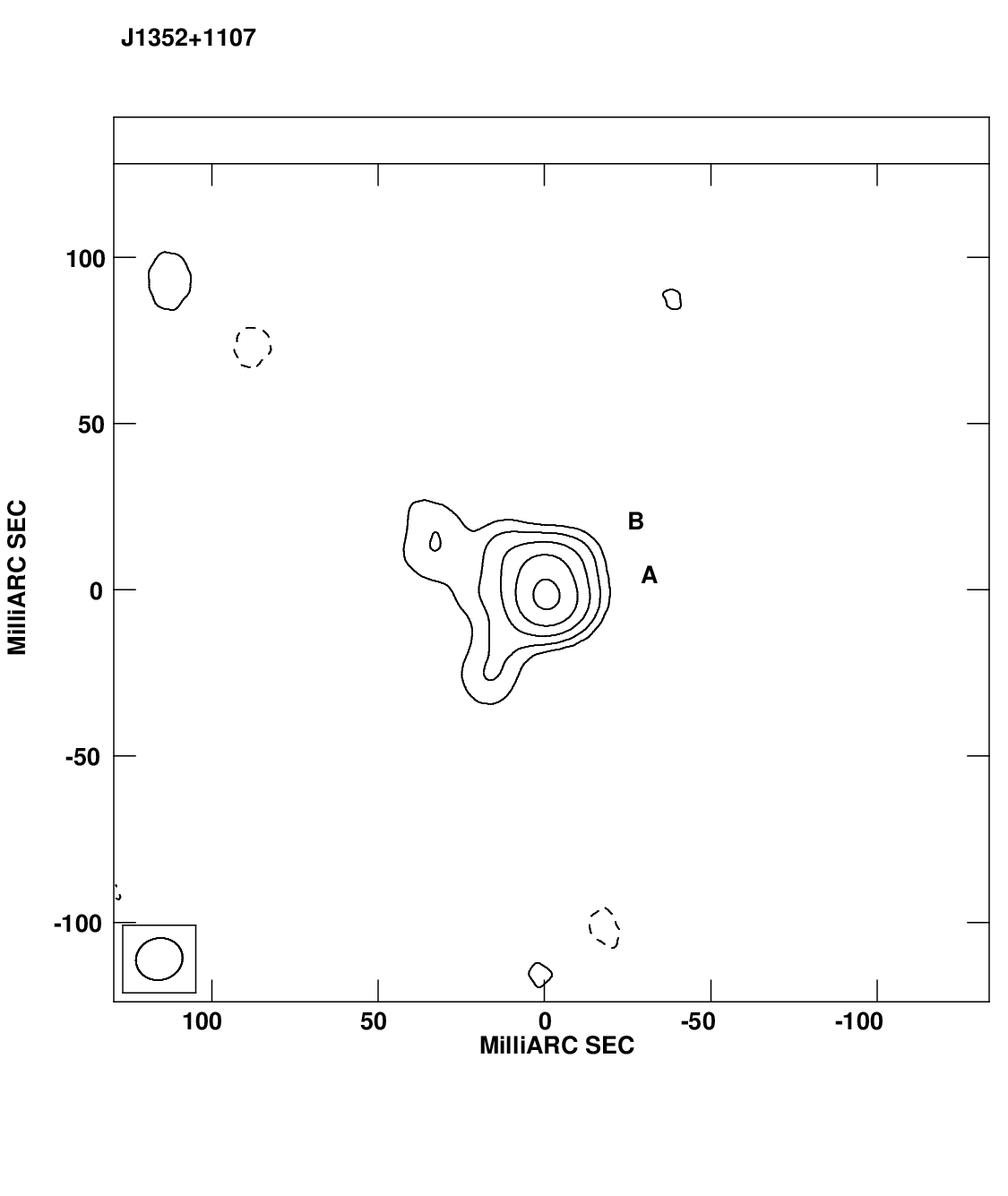}
     \caption{J1352+1107 at 1.65 GHz, the restoring beam is
        $14.0\times12.5$ mas with PA $-75.3^{\circ}$, the peak is 379 mJy/beam, the first contour
        is 20 mJy/beam.}
      \label{fig10}
   \end{figure}

The 1.6 GHz VLBI image (Fig.~\ref{fig10}) is the first VLBI image
of the source. It appears to have a compact double structure or a
core-jet alike, and seems diffuse emission around the source.
About 30\% total flux density (estimated from
Table~\ref{table:gps}) is resolved out in the VLBI image. Either a
core-jet or a compact double classification is possible.

\subsection{J1600$-$0037 (PKS B1557$-$004)}

\begin{figure}
     \includegraphics[width=7cm]{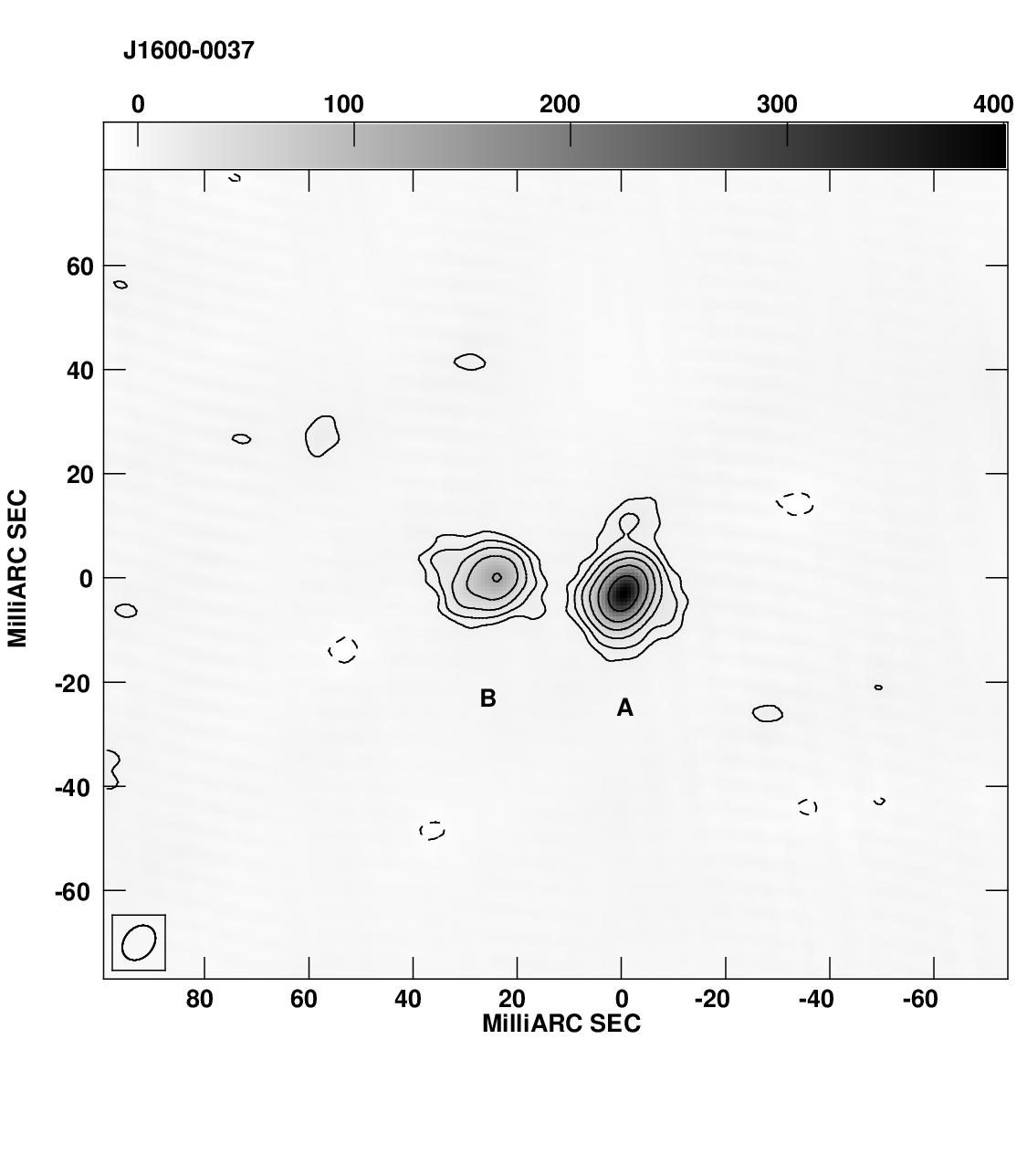}
     \caption{J1600$-$0037 at 1.65 GHz, the restoring beam is
        $7.4\times5.5$ mas with PA $-39.8^{\circ}$, the peak is 400 mJy/beam, the first contour
        is 8 mJy/beam.}
      \label{fig11}
   \end{figure}

The 1.6 GHz VLBI image (Fig.~\ref{fig11}) is the first VLBI image
of the source, and it has an overall double structure, the eastern
component has some extension in the west-east direction. A flux
variability of $(13\pm6.9)\%$ at 5 GHz in Table~\ref{table:flux}
may suggest this is a core-jet source.

\subsection{J1648+0242 (4C+02.43)}

The GPS source is not detected with VLBI. It is an NVSS
double-lobe source, and totally resolved out in the VLBI
observation.

\subsection{J2058+0540 (4C+05.78)}

\begin{figure}
     \includegraphics[width=7cm]{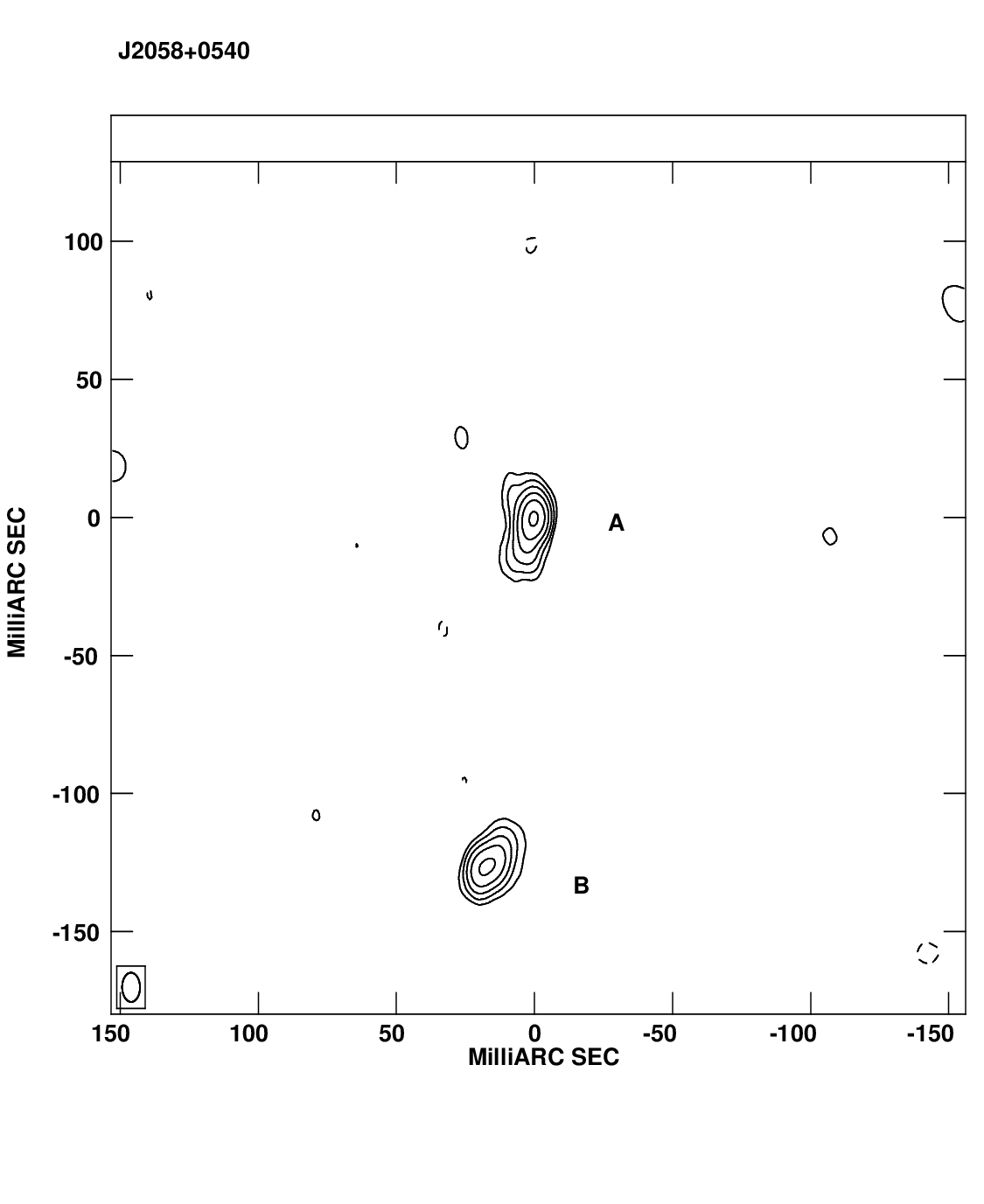}
     \caption{J2058+0540 at 1.65 GHz, the restoring beam is
        $10.6\times6.5$ mas with PA $0.7^{\circ}$, the peak is 362 mJy/beam, the first contour
        is 10 mJy/beam.}
      \label{fig12}
   \end{figure}

The 1.6 GHz VLBI image (Fig.~\ref{fig12}) is the first VLBI image
of the source. It shows a double-lobe source, and for the size of
970 pc, we suggest this is a CSO.

\subsection{PKS B2121$-$014}

\begin{figure}
     \includegraphics[width=7cm]{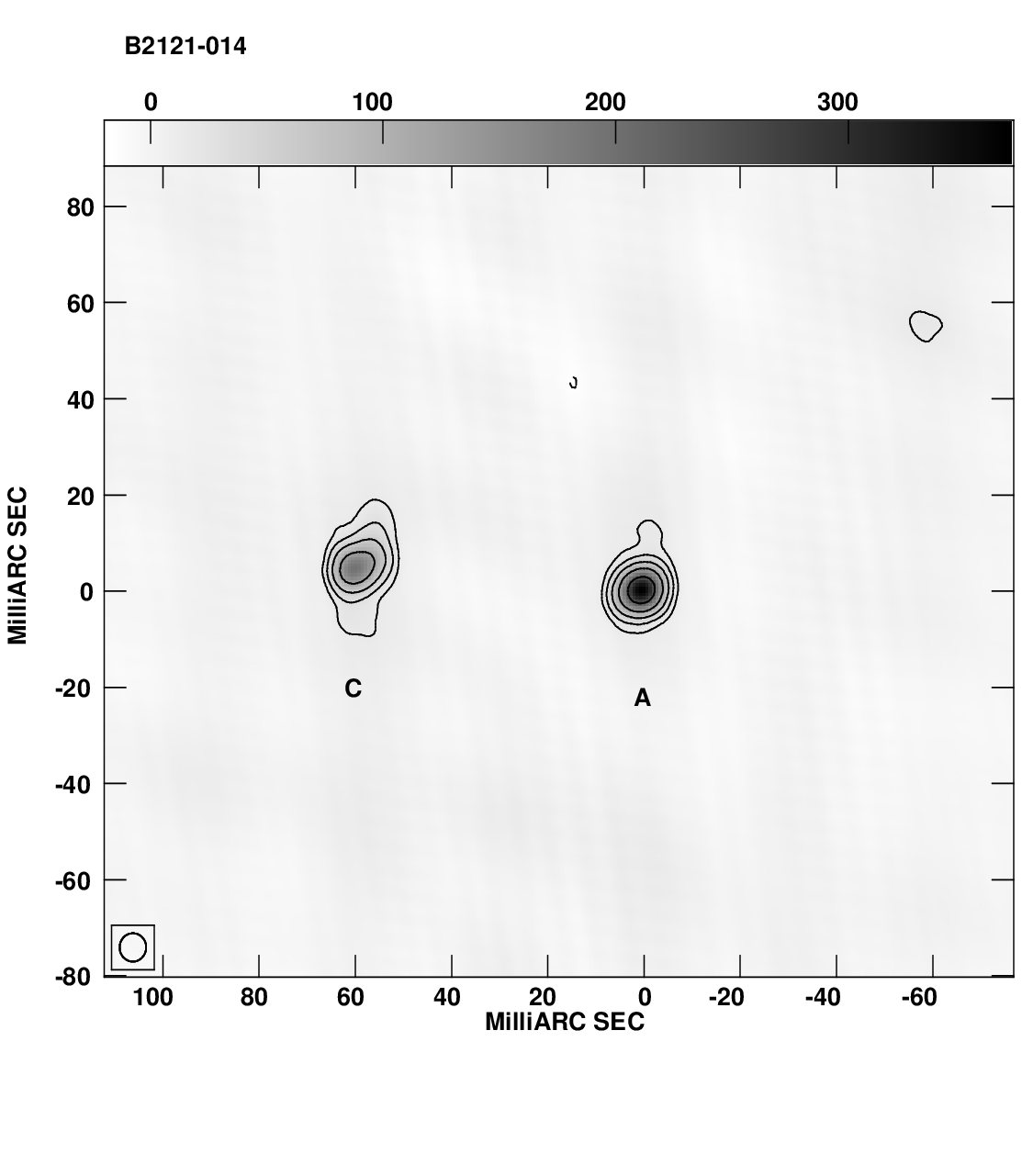}
     \caption{2121$-$014 at 1.65 GHz, the restoring beam is
        $5.9\times5.5$ mas with PA $-1.4^{\circ}$, the peak is 370 mJy/beam, the first contour
        is 15 mJy/beam.}
      \label{fig13}
   \end{figure}

The 1.6 GHz VLBI image (Fig.~\ref{fig13}) shows a double-lobe
structure, it is similar to that at 2.3 and 5 GHz (Xiang et al.
2005, 2006), except that a weak jet-like emission `B', which
appears at 2.3 and 5 GHz, is missing, probably due to absorption
at the lower frequency 1.6 GHz. The source is a CSO for the source
size of 488 pc.

\subsection{PKS B2322$-$040}

\begin{figure}
     \includegraphics[width=7cm]{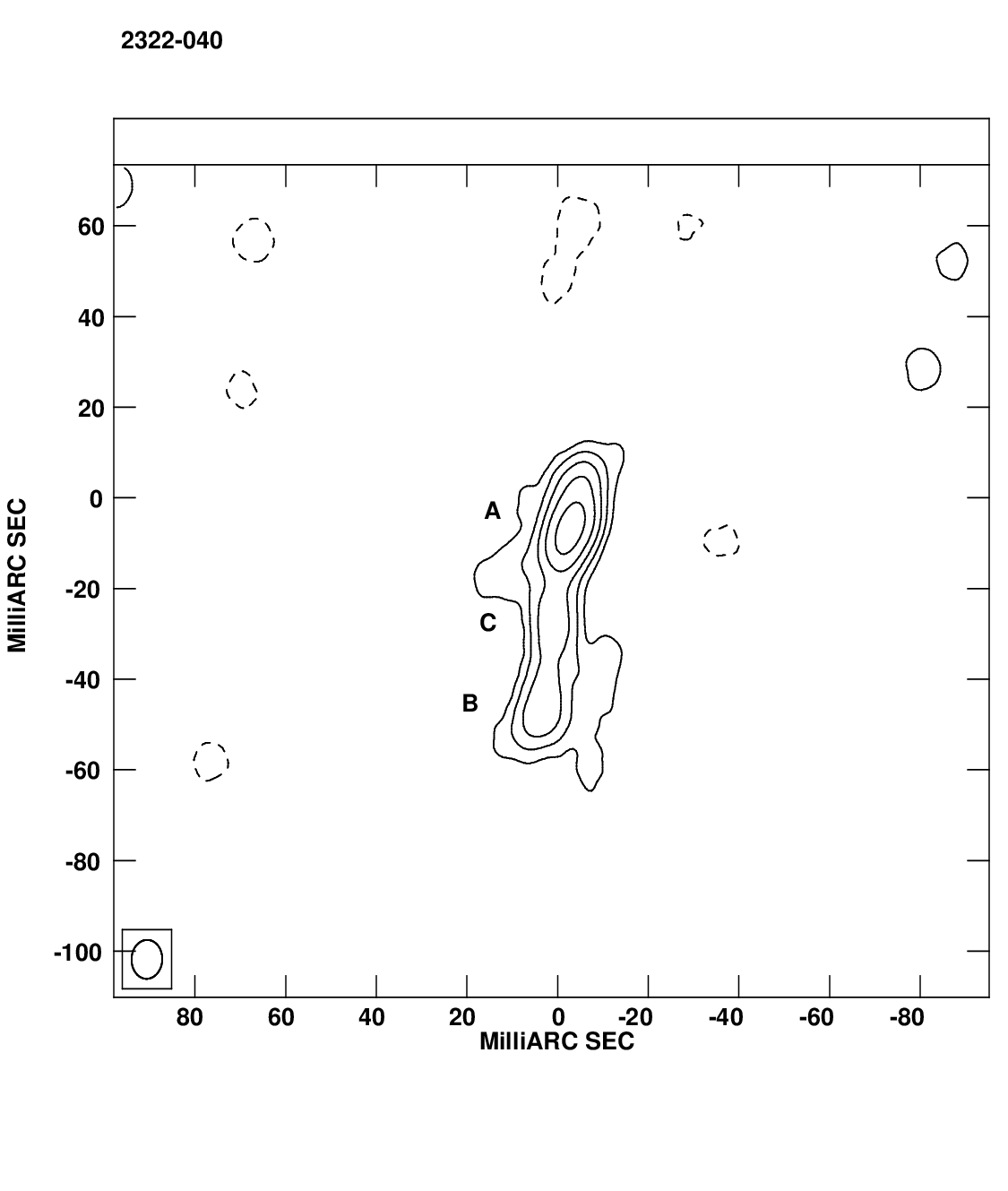}
     \caption{2322$-$040 at 1.65 GHz, the restoring beam is
        $8.6\times6.8$ mas with PA $-2.1^{\circ}$, the peak is 233 mJy/beam, the first contour
        is 10 mJy/beam.}
      \label{fig14}
   \end{figure}

The 1.6~GHz VLBI image (Fig.~\ref{fig14}) exposes a central
emission region between the two lobes `A' and `B', which is
probably a core embedded in the central region. The `core'
emission is not detected at higher frequencies (Xiang et al. 2005,
2006), but it emerges at 1.6 GHz near the peak frequency (1.4 GHz)
of the GPS source. There is a flux increase of $(4.0\pm1.9)\%$
over 15 years at 5 GHz (Table~\ref{table:flux}), may suggest that
the core is currently active. The source can be a CSO for its size
of 450 pc.

\subsection{PKS B0914+114}

\begin{figure}
     \includegraphics[width=7cm]{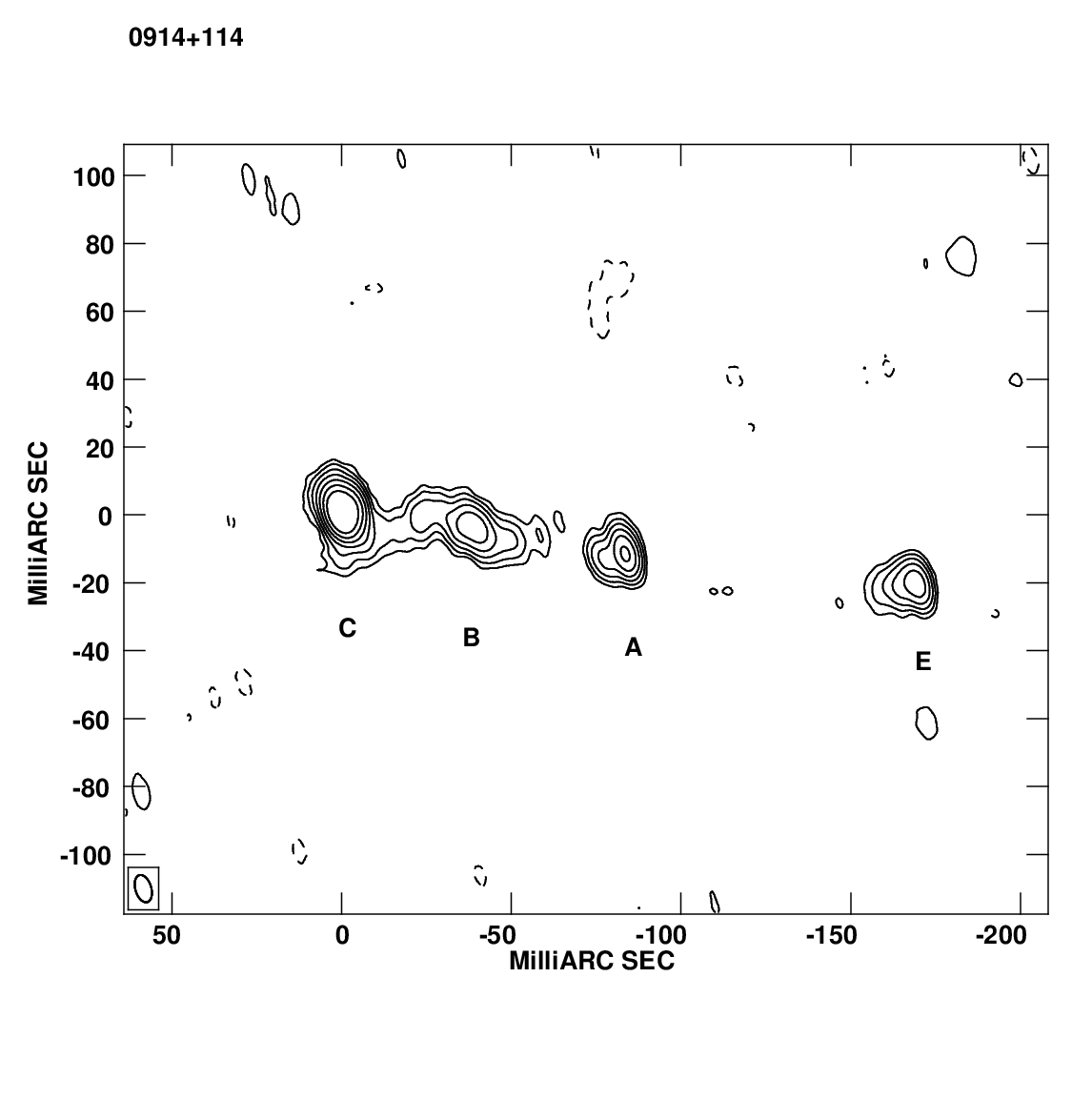}
     \caption{0914+114 at 1.65 GHz, the restoring beam is
        $8.3\times4.7$ mas with PA $16.8^{\circ}$, the peak is 246 mJy/beam, the first contour
        is 1 mJy/beam.}
      \label{fig15}
   \end{figure}

 The 1.6 GHz VLBI image (Fig.~\ref{fig15}) exhibits a core `A', jet feature `B' and two lobes `C', `E'.
 The western one `E'
 emerges at this frequency. Labiano et al. (2007)
 have identified an empty field ($> 25\, m_{R}$) at the FIRST position
 of the source, and concluded that the previously identified nearby disk galaxy (a redshift
 of 0.178) is not the host of this radio source 0914+114. For the typical compact symmetric structure,
 we consider the source is a CSO.

\subsection{1751+278 (MG2 J175301+2750)}

\begin{figure}
     \includegraphics[width=7cm]{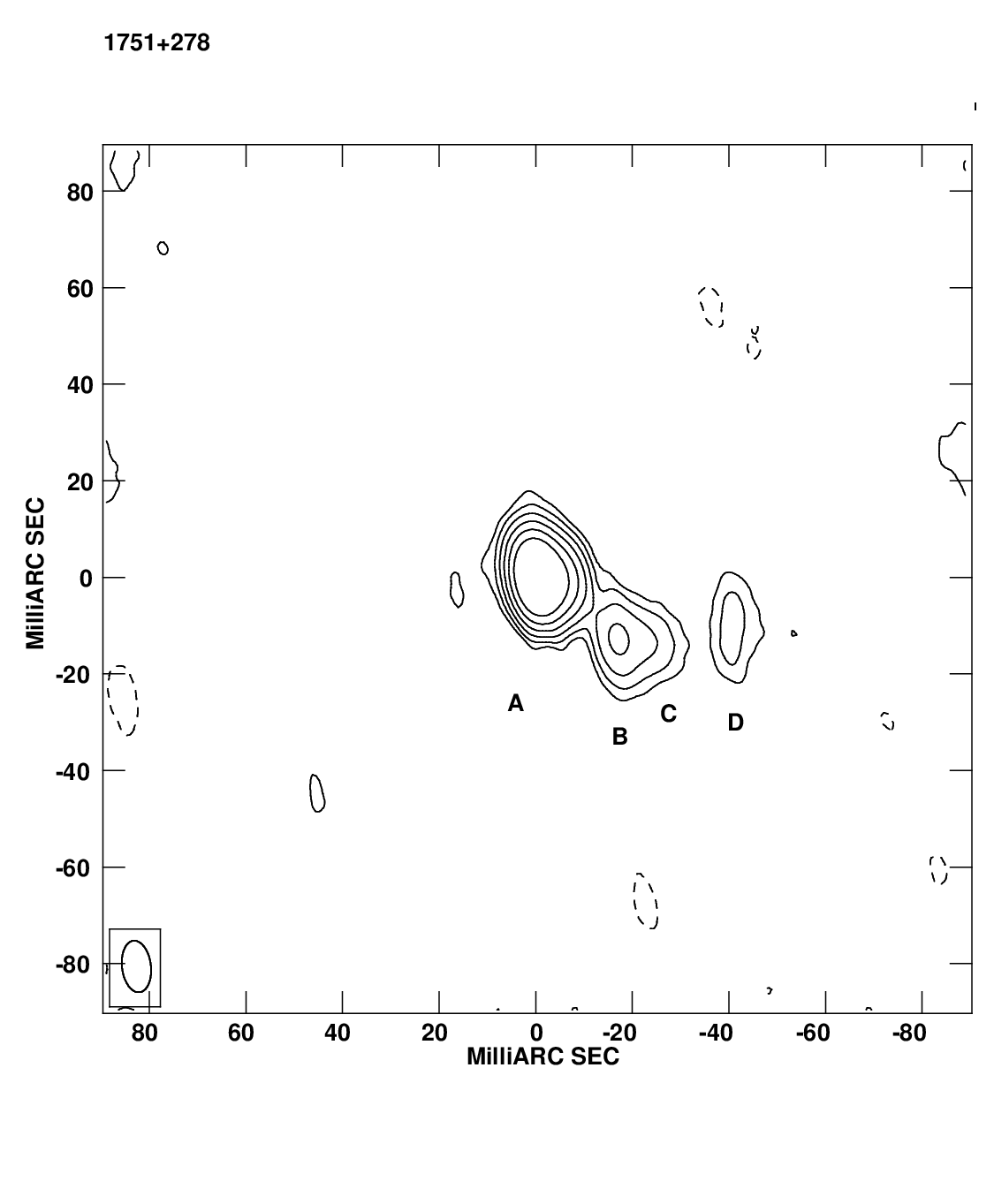}
     \caption{1751+278 at 1.65 GHz, the restoring beam is
        $10.7\times6.0$ mas with PA $7.0^{\circ}$, the peak is 406 mJy/beam, the first contour
        is 3 mJy/beam.}
      \label{fig16}
   \end{figure}

The 1.6 GHz structure (Fig.~\ref{fig16}) is similar to what we got
before at 1.6 GHz (Xiang et al. 2002), and confirms that there is
jet-like emission `C' and `D' associated with the southern
component `B', indicating this is a core-jet source.

\subsection{B2 1824+271 }

\begin{figure}
     \includegraphics[width=7cm]{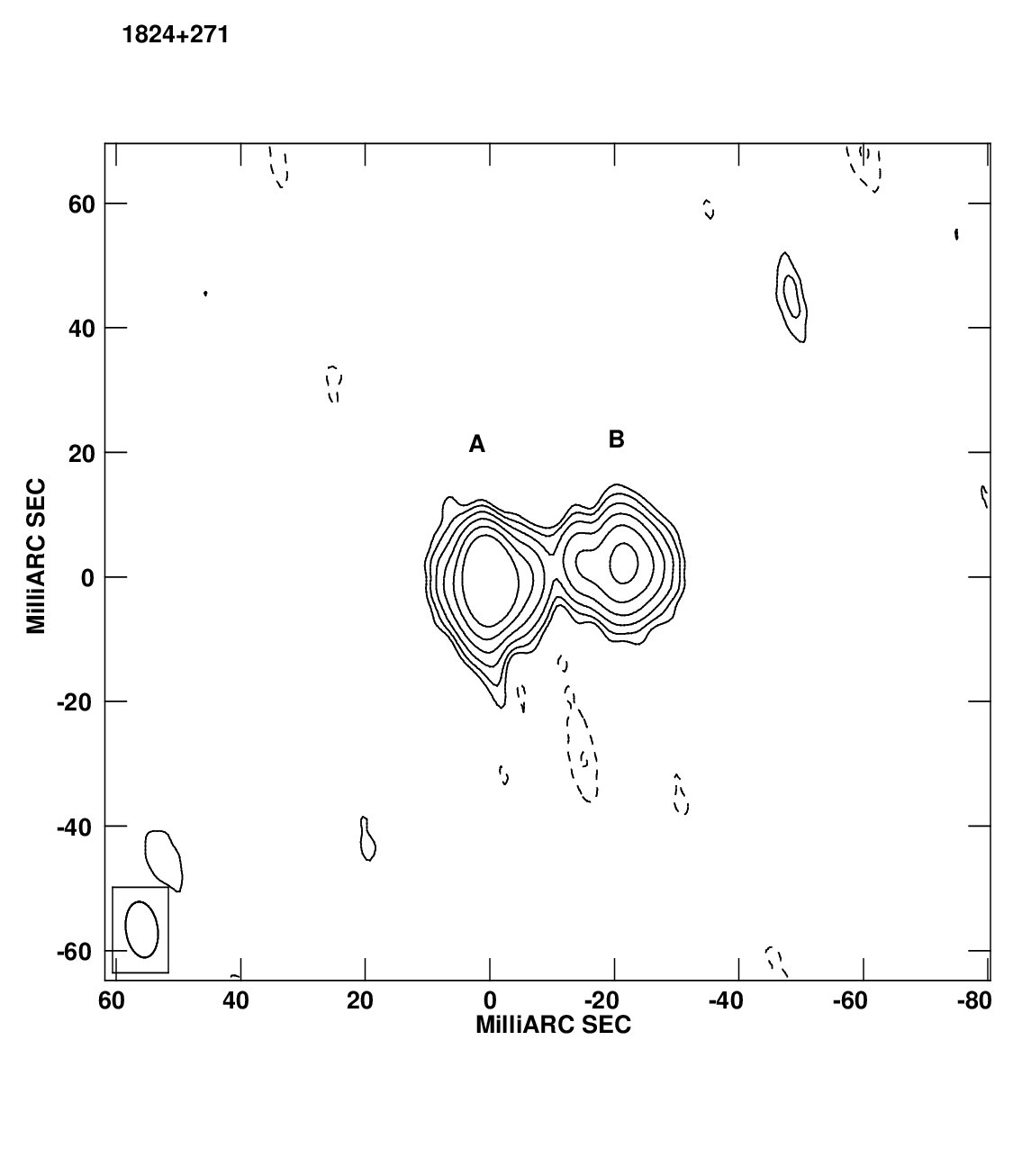}
     \caption{1824+271 at 1.65 GHz, the restoring beam is
        $9.0\times5.1$ mas with PA $8.1^{\circ}$, the peak is 146 mJy/beam, the first contour
        is 1 mJy/beam.}
      \label{fig17}
   \end{figure}

 The 1.6 GHz VLBI image (Fig.~\ref{fig17}) exposes a
symmetric double structure and jet-like emission associated with
the two lobes, confirming this is a CSO as we have suggested
(Xiang et al. 2006).

\subsection{[WB92] 2323+790}

\begin{figure}
     \includegraphics[width=7cm]{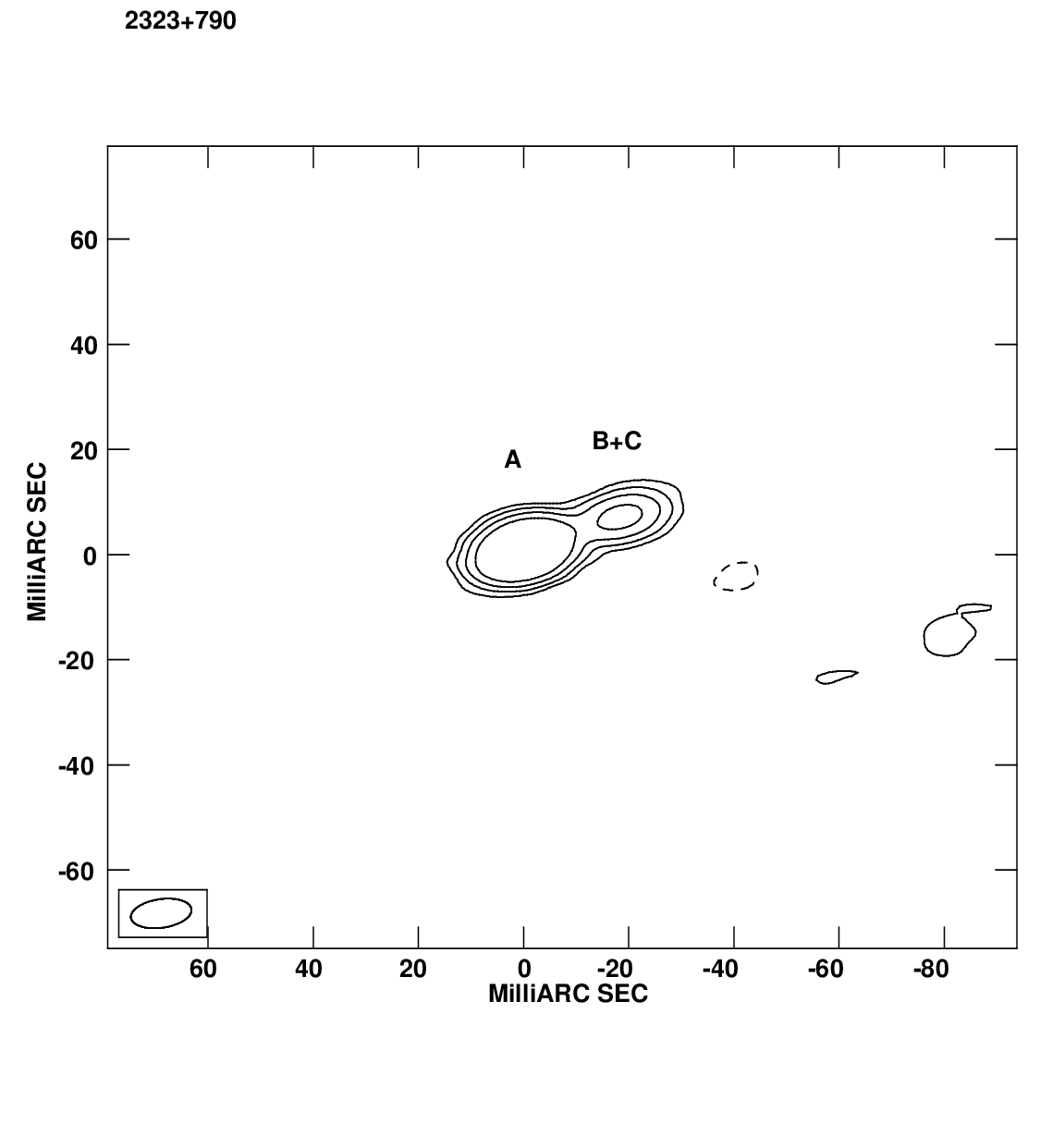}
     \caption{2323+790 at 1.65 GHz, the restoring beam is
        $11.6\times5.4$ mas with PA $-82^{\circ}$, the peak is 438 mJy/beam, the first contour
        is 10 mJy/beam.}
      \label{fig18}
   \end{figure}

 The 1.6 GHz image (Fig.~\ref{fig18}) shows a central component `A' and a weak one `B+C' in the north-west,
 and the components `A' and `B+C' show steep spectra between 1.6 GHz and 5 GHz (Xiang et al.
 2006). The source can be a CSO candidate.

\section{Discussion}

In the sample (Table~\ref{table:gps}), J1648+0242 is an NVSS
double source and is not detected in this VLBI observation; all
others are point-like in the NVSS images, indicating that GPS
sources are compact. Except four sources (J1057+0012, J1352+1107,
J1600$-$0037 and 1751+278), 14 out of 18 sources exhibit double or
triple VLBI structure and can be CSOs or CSO candidates though
some of them have no measured redshift. The sources with redshift
show double or triple structure with sizes $<1$ kpc, suggesting
these GPS sources are certainly compact and likely CSOs.

The mini double-lobe sources or CSOs could be more stable in flux
density than other type of compact sources. We have measured the
flux densities for the sources (Table~\ref{table:flux}) at 4.85
GHz and compared with the values observed 15 years ago, we found
that 12 among 14 GPS sources are likely stable in flux ($1\sigma$
level), two sources (J1057+0012 and J1600$-$0037) show about 10\%
variability in $3\sigma$ and $2\sigma$ level respectively. The
flux variability on J1057+0012 and J1600$-$0037 is consistent with
their core-jet classification. `Core-jet' sources are defined to
show a one-sided jet, and the jet is often closely pointing to us
(from a pole-on AGN). It is hard to estimate the real source size
due to Doppler boosting, hence the `core-jet' sources might not be
young radio sources even if they appear to be compact in some
cases.

In addition, some sources are resolved out in our VLBI image by
more than 10\% of total flux estimated from Table~\ref{table:gps},
probably due to diffuse emission associated with lobes and
tail/jet emission. They are J0210+0419 (-14\%), J0323+0534
(-38\%), J0433$-$0229 (-18\%),
 J1352+0232 (-15\%), J1135+1107 (-31\%),
J2058+0540 (-12\%), 2322$-$040 (-15\%), and J1648+0242 is
completely resolved out. The VLBI flux densities of the other nine
sources at 1.6 GHz are consistent with the estimated total flux
densities within an error of 10\% the estimated amplitude
uncertainty of the EVN observations.

\section{Summary and conclusion}

\begin{enumerate}

\item We obtained total intensity 1.6~GHz VLBI images of 17 GPS
sources for the first time. The majority (80\%) show
mini-double-lobe radio structure, indicating that they are CSOs or
candidates, and their host AGNs could be edge-on to us. This
result suggests that there is a high incidence of mini double-lobe
sources and CSOs in the GPS source sample.

\item The sources J0323+0534, J1135$-$0021, J1352+0232,
J2058+0540, 2121$-$014 and 2322$-$040 with measured redshift, are
double-lobed with sizes of $<1$ kpc, and are classified as CSOs.

\item Three sources (J1057+0012, J1600$-$0037 and 1751+278) are
classified as core-jet sources according to their morphologies and
flux variability.

\item The 1.6 GHz images of the sources 0914+114, 1824+271,
2121$-$014 and 2322$-$040, for which we had observations at 2.3,
5.0 and 8.4 GHz, have provided information on their source
structure and spectra at the lower frequency, permitting further
spectral study in the future.

\end{enumerate}

\begin{acknowledgements}
We thank the referee Alvaro Labiano, and Nathan de Vries for
comments. The European VLBI Network is a joint facility of
European, Chinese, South African and other radio astronomy
institutes funded by their national research councils. This
research has made use of the NASA/IPAC Extragalatic Database (NED)
which is operated by the Jet Propulsion Laboratory, Caltech, under
contract with NASA. This work was partly supported by the Natural
Science Foundation of China (NSFC).
\end{acknowledgements}

\clearpage


\clearpage

\clearpage

\clearpage


\begin{thebibliography}{}


\bibitem[]{}
de Vries N., Snellen I. A. G., Schilizzi R. T., Lehnert M. D.,
Bremer M. N., 2007, A\&A 464, 879

\bibitem[]{}
de Vries W. H., Barthel P. D., O'Dea C. P., 1997, A\&A 321, 105

\bibitem[]{}
Dallacasa D., 2004, in Proceedings of the 7th EVN Symposium, eds:
Bachiller R., Colomer F., et al.

\bibitem[]{}
Fanti, C., Fanti, R., Dallacasa, D., Schilizzi, R. T., Spencer, R.
E., Stanghellini, C. 1995, A\&A 302, 317


\bibitem[]{}
Gregory P. C., Condon J. J., 1991, ApJS 75, 1011

\bibitem[]{}
Griffith M. R., Wright A. E., Burke B. F., Ekers R. D., 1995, ApJS
97, 347




\bibitem[]{}
Labiano A, Barthel P. D., O'Dea C. P., de Vries W. H., Perez I.,
Baum S. A., 2007, A\&A 463, 97L

\bibitem[]{}
Murgia M., 2003, PASA 20, 19

\bibitem[]{}
O'Dea C. P. 1998, PASP 110, 493

\bibitem[]{}
Orienti M., Dallacasa D., Stanghellini C., 2007, A\&A 461, 923

\bibitem[]{}
Polatidis A. G., Conway J. E., 2003, PASA 20, 69


\bibitem[]{}
Owsianik I., Conway J. E., 1998, A\&A 337, 69






\bibitem[]{}
Snellen I. A. G., Lehnert M. D., Bremer M. N., Schilizzi R. T.,
2002, MNRAS 337, 981

\bibitem[]{}
Snellen I. A. G., Schilizzi R. T., Miley G. K., de Bruyn A. G.,
Bremer M. N., R\"ottgering H. J. A., 2000, MNRAS 319, 445

\bibitem[]{}
Tschager W., Schilizzi R. T., R\"ottgering H. J. A., Snellen I. A.
G., Miley G. K., 2000, A\&A 360, 887

\bibitem[]{}
Stanghellini C., O'Dea C. P., Dallacasa D., Cassaro P., Baum S.
A., Fanti R., Fanti C., 2005, A\&A 443, 891

\bibitem[]{}
Stanghellini C., O'Dea C. P., Dallacasa D., Baum S. A., Fanti R.,
Fanti C., 1998, A\&AS 131, 303

\bibitem[]{}
Tinti S.,Dallacasa D., de Zotti G., Celotti A., Stanghellini C.,
2005, A\&A 432, 31

\bibitem[]{}
Xiang L., Stanghellini C., Dallacasa D., Haiyan Z., 2002, A\&A
385, 768

\bibitem[]{}
Xiang L., Dallacasa D., Cassaro P., Jiang D., Reynolds C., 2005,
A\&A 434, 123

\bibitem[]{}
Xiang L., Reynolds C., Strom R. G., Dallacasa D., 2006, A\&A 454,
729

\end{thebibliography}
\end{document}